\documentclass[showpacs,aps,prd,amsmath,amssymb,nofootinbib,superscriptaddress]{revtex4-2}
\usepackage{graphicx}
\usepackage{amssymb}
\usepackage{bbm}
\usepackage{hyperref}
\usepackage{slashed}
\usepackage{float}
\usepackage{subfigure}
\usepackage{dcolumn}

\usepackage{multirow}
\usepackage{color}
\begin{document}

\title{Polarized and unpolarized off-shell $H^\ast\to ZZ\rightarrow 4\ell$ decay above the $2m_Z$ threshold}
\author{A. I. Hern\'andez-Ju\'arez}
\email{alan.hernandez@cuautitlan.unam.mx}
\affiliation{Departamento de F\'isica, FES-Cuautitl\'an, Universidad Nacional Aut\'onoma de M\'exico, C.P. 54770, Estado de M\'exico, M\'exico.}
\author{R. Gait\'an}
\affiliation{Departamento de F\'isica, FES-Cuautitl\'an, Universidad Nacional Aut\'onoma de M\'exico, C.P. 54770, Estado de M\'exico, M\'exico.}
\author{G. Tavares-Velasco}
\affiliation{Facultad de Ciencias F\'isico Matem\'aticas, Benem\'erita Universidad Aut\'onoma de Puebla, Apartado Postal 1152, Puebla, Puebla, M\'exico}

\begin{abstract}
An analysis of the off-shell $H^\ast\rightarrow ZZ \rightarrow \overline{\ell}_1\ell_1\overline{\ell}_2\ell_2$ decay width is presented for both unpolarized and polarized $Z$ gauge bosons in the scenario with the most general $H^*ZZ$ vertex function, which is given in terms of two $CP$-even ($\hat b_Z$ and $\hat c_Z$) and one $CP$-odd  ($\tilde b_Z$) anomalous couplings.
 The SM contributions to the $H^*ZZ$ coupling up to the one-loop level are also included. Explicit analytic results for the unpolarized and polarized $H^\ast\rightarrow ZZ \rightarrow \overline{\ell}_1\ell_1\overline{\ell}_2\ell_2$  square amplitudes and the four-body phase space are presented, out of which several observable quantities can be obtained straightforwardly.  As far as the numerical analysis is concerned, a cross-check  is performed  via \texttt{MadGraph5\_aMC@NLO}, where our model was implemented with the aid of FeynRules.
We then  consider the most stringent bounds on anomalous complex $H^*ZZ$ couplings and analyze the effects of the polarizations of the $Z$ gauge bosons through the polarized $H^\ast\rightarrow ZZ \rightarrow \overline{\ell}_1\ell_1\overline{\ell}_2\ell_2$ decay width as well as left-right and forward-backward asymmetries, which are found to be sensitive to new-physics effects. Particular focus is put on  the effects of the absorptive parts of the anomalous $H^*ZZ$ couplings, which have been largely overlooked up to now in LHC analyses.  It is found that the studied observable quantities, particularly the left-right asymmetries, can be helpful to look for effects of $CP$-violation in the $H^*ZZ$ coupling and set bounds on the absorptive parts. For completeness, we also analyze the case of unpolarized $Z$ gauge bosons.

\end{abstract}
\date{\today}

\maketitle
\section{Introduction}
\label{intro}

All the experimental data collected until now indicate that the properties of the scalar particle discovered in 2012 at the LHC by the ATLAS and CMS collaborations \cite{CMS:2012qbp,ATLAS:2012yve} are compatible with those of the long-sought-for Higgs boson, which is the remanent of the mechanism of spontaneous symmetry breaking of the $SU(2)_L\times U(1)_Y$ gauge symmetry in the Glashow-Salam-Weinberg Standard Model (SM). Nonetheless, there are still several couplings of such a particle that remain to be measured with enough accuracy, such as for the $H\bar{b}b$ and $H\mu^-\mu^+$ interactions, whereas other ones are still not yet at the reach of experimental measurement, for instance, the Higgs boson couplings to light fermions, let alone the Higgs boson self couplings.  Therefore, the study of the Higgs boson phenomenology is expected to play an important role in probing the SM and the search for any new physics effects in present and future colliders. Along these lines, the Higgs boson couplings to the weak gauge bosons $HZZ$ and $HWW$, which in the SM arise at the tree level, have long drawn considerable attention both theoretically and experimentally. 
As far as the $HZZ$ coupling is concerned, it has played a crucial part in the experimental study of the Higgs boson at the LHC via the $H\to ZZ^*\to 4\ell$ decay, which provides a clear signal and allows for a precise resolution of the Higgs boson mass despite that its  branching ratio is considerably smaller than those of other decay channels such as $H\to WW^*$, $H\to \tau^+\tau^-$, and $H\to\bar{b}b$. A milestone  was achieved in 2022 when the CMS  collaboration reported for the first time the measurement of the off-shell $H^*ZZ$ coupling via the $pp\to H^*\to ZZ$ process \cite{CMS:2022ley}. This result was also confirmed by the ATLAS collaboration the following year \cite{ATLAS:2023dnm}. 

Although one-loop corrections to the $HZZ$ vertex  were calculated long ago in the framework of the SM \cite{Kniehl:1990mq,Phan:2022amy}, more recently a  new evaluation was presented \cite{Hernandez-Juarez:2023dor} that could be more suitable for the recent progress on the theoretical and experimental study of this coupling.  

The most general $HZZ$ vertex can be parametrized by an effective Lagrangian given in terms of four coupling constants:
\begin{align}
\label{Lag}
\mathcal{L}^{HZZ}=&\frac{g }{c_W}m_Z \left[\frac{(1+a_Z)}{2}  H Z_\mu Z^\mu+\frac{1}{2m^2_Z} \Big\{\hat{b}_Z HZ_{\mu\nu}Z^{\mu\nu}+\hat{c}_Z HZ_\mu\partial_\nu Z^{\mu\nu}+ \widetilde{b}_Z H Z_{\mu\nu}\widetilde{Z}^{\mu\nu}\Big\}\right],
\end{align}
where  $a_Z$ stands for the  corrections to the SM tree-level coupling via loop contributions or new physics effects, whereas $\hat{b}_Z$, $\hat{c}_Z$ and  $\widetilde{b}_Z$ are absent at the tree level in the SM and thus are dubbed anomalous couplings. The  anomalous $CP$-conserving coupling  $\hat{b}_Z$ is induced at the one-loop level \cite{Kniehl:1990mq,Phan:2022amy} and can reach values of the order of $10^{-3}$ \cite{Hernandez-Juarez:2023dor}, while the $CP$-violating one $\widetilde{b}_Z$ would arise up to the three-loop level and has been estimated to be of the order of $10^{-11}$ \cite{Soni:1993jc}. The $\hat{c}_Z$ anomalous coupling is also expected to be generated at one-loop level, but it has not been identified in SM calculations.

The phenomenology of the $HZZ$ anomalous couplings have been extensively studied in both lepton \cite{Hagiwara:2000tk, Biswal:2005fh, Dutta:2008bh} and hadron colliders \cite{Javurkova:2024bwa}. These studies have shown that non-SM contributions to the $HZZ$ coupling could be significantly constrained in the future, providing an opportunity to detect new physics effects through $HZZ$-mediated processes.  In particular, polarization observables in the $Z^\ast\rightarrow HZ$ process at the LHC have been used to revisit the anomalous couplings, yielding limits of around $10^{-3}-10^{-4}$ for the real and absorptive parts of the $CP$-conserving $\hat{b}_Z$ anomalous coupling and of order $10^{-3}$ for the $CP$-violating $\widetilde{b}_Z$ coupling \cite{Rao:2020hel}. Furthermore, limits on the anomalous couplings have been established at various colliders, including $e^-e^-$, $ep$, and $\gamma e$ colliders  \cite{Kniehl:2001jy, Biswal:2005fh, Godbole:2007cn, Rao:2023ogi, Fabbrichesi:2023jep, Cakir:2013bxa, Sahin:2019wew, Gauld:2023gtb, Klein:2023kuy, Rao:2023jpr, Xiong:2023qnr, Asteriadis:2024qim}. The CMS collaboration has also obtained bounds on effective ratios of the anomalous couplings \cite{CMS:2022ley, CMS:2017len, CMS:2019ekd, CMS:2021nnc}, which were combined with the theoretical SM contribution to $\hat{b}_Z$ in a recent study  \cite{Hernandez-Juarez:2023dor}. It was found that the $\hat{c}_Z$ and $\widetilde{b}_Z$ anomalous couplings can be constrained to as tight as $10^{-2}-10^{-4}$ depending on the energy region. Finally, other off-shell couplings such as trilinear neutral gauge boson couplings \cite{Gounaris:2000tb, Choudhury:2000bw, Hernandez-Juarez:2021mhi, Hernandez-Juarez:2022kjx} and the coupling of the gluon with a quark-antiquark pair \cite{Hernandez-Juarez:2020drn, Hernandez-Juarez:2020gxp} have also been investigated in recent studies.

In this work, we are interested in the study of  the $HZZ$ anomalous couplings effects on the off-shell decay $H^*\to ZZ\to 4\ell$ via polarized $Z$ gauge bosons. The polarizations of $Z$ boson  are of great interest at the LHC, as recently a longitudinally polarized $Z$ boson pair has been reported by the ATLAS collaboration \cite{ATLAS:2023zrv}, and the production of longitudinally $W^\pm Z$ pairs is also being studied \cite{ATLAS:2022oge}. The polarization fractions of the $Z$ boson have been measured by the  CMS, ATLAS and LHCb collaborations \cite{CMS:2015cyj,ATLAS:2016rnf,LHCb:2022tbc,ATLAS:2023lsr}. Previous analysis  of gauge bosons polarization at the LHC include $W^\pm Z$ production  \cite{ATLAS:2019bsc}, $W$+Jets events \cite{CMS:2011kaj,ATLAS:2012au}, $W$ bosons produced in top decays \cite{ATLAS:2016fbc,CMS:2016asd,CMS:2020ezf,ATLAS:2022rms}, $W^\pm W^\pm$ production \cite{CMS:2020etf}. The \texttt{MadGraph5\_aMC@NLO} \cite{BuarqueFranzosi:2019boy} and \texttt{SHERPA} \cite{Hoppe:2023uux} event generators have also included the possibility of generating polarized amplitudes.  A method to identify $Z$ bosons polarizations has been discussed in Ref. \cite{Ballestrero:2019qoy}, whereas a higher sensitivity to polarized $Z$ bosons is also expected in the recent upgrade of the LHC \cite{CMS:2018mbt}. We will consider the most general scenario with both $Z$ gauge bosons on-shell and complex anomalous $HZZ$ couplings. Furthermore, we will analyze the behavior of  some observables that can be sensitive to new physics contributions. In particular, we are interested in the effects of the absorptive parts of the anomalous $HZZ$ couplings, which have been overlooked in the past but can lead to interesting results. For the sake of completeness, we also discuss the scenario with unpolarized $Z$ gauge bosons.

The rest of this work is as follows. Section \ref{framework} is devoted to discuss the theoretical framework appropriate for the study of the $H^*\to ZZ\to 4\ell$ decay width, with explicit analytical expressions for the polarized and unpolarized square amplitudes, out of which the respective  $H^*\to ZZ\to 4\ell$ differential decay widths are obtained. The numerical analysis is presented in Sec. \ref{numanal}, where we cross-check  our calculation method with an alternative evaluation performed via \texttt{MadGraph5\_aMC@NLO}. The behavior of the differential decay, angular distributions along with  left-right and a forward-backward asymmetries are analyzed in some realistic scenarios for the values of the real and absorptive parts of the anomalous $HZZ$ couplings. In Section \ref{conclusions} we present our concluding remarks and outlook. Finally, the  kinematics of the $H^*\to ZZ\to 4\ell$ decay and the four-body phase space are discussed in Appendix \ref{kinematics}.

\section{Polarized and unpolarized $H^\ast\rightarrow \overline{\ell}_1\ell_1\overline{\ell}_2\ell_2$ decay widths}
\label{framework}
We now present all the analytical formulas required for the calculation of the unpolarized and polarized $H^\ast\rightarrow \overline{\ell}_1\ell_1\overline{\ell}_2\ell_2$ decay width. The most general form of the $HZZ$ vertex functions with complex anomalous couplings is considered, whereas   the contributions of the SM up to the one-loop level are also included \cite{Hernandez-Juarez:2023dor}.

\subsection{$HZZ$ vertex function}

The $S$-matrix element obtained from Lagrangian \eqref{Lag} can be written as
\begin{equation}
i\mathcal{M}=i\frac{g}{c_W}m_Z \Gamma^{ZZH}_{\mu\nu}(q^2) Z^\mu(p_1) Z^\nu(p_2) H(q),
\end{equation}
where the vertex function $\Gamma^{ZZH}_{\mu\nu}(q^2)$ can be written as follows for on-shell $Z$ gauge bosons and an off-shell Higgs boson
\begin{align}
\label{vertex}
\Gamma_{\mu\nu}^{ZZH}(q^2)=h^H_1(q^2) g_{\mu\nu}+\frac{h_2^H(q^2)}{m_Z^2} p_{1\nu}p_{2\mu}+\frac{h_3^H(q^2) }{m_Z^2}\epsilon_{\mu\nu\alpha\beta}p_{1}^\alpha p_{2}^\beta,
\end{align}
with $h_i^H$  ($i=$1, 2, 3) the form factors that are given in terms of the anomalous couplings of  Lagrangian \eqref{Lag} as follows
\begin{align}
& h_1^H(q^2)=1+ a_Z-   \hat{b}_Z \frac{q^2-2m_Z^2}{m_Z^2}+ \hat{c}_Z, \label{H11}\\
&h_2^H(q^2)= 2 \hat{b}_Z, \label{H22}\\
&h_3^H(q^2)=  2 \widetilde{b}_Z. \label{H33}
\end{align}
These form factors are both explicit and implicit functions of $q^2$ since  the anomalous couplings also depend on the off-shell Higgs boson four-momentum. 

As the main interest of this work is the study of the anomalous contributions, we will consider a vanishing $a_Z$. For a study of the SM one-loop contributions to the $a_Z$ coupling along with the renormalization procedure,  we refer the reader to Ref. \cite{Kniehl:1990mq}.  The $\hat{b}_Z$, $\hat{c}_Z$ and $\widetilde{b}_Z$ anomalous couplings will be considered as non-zero in the rest of this work. This is feasible in the SM, and new physics contributions from new particles can also arise under the context of the two-Higgs doublet and Higgs singlet models \cite{Kanemura:2017wtm}, the minimal Higgs triplet model \cite{Aoki:2012jj} and the minimal supersymmetric SM \cite{Englert:2014ffa}. Recently, the SMEFT contributions have also been  revisited \cite{Asteriadis:2024qim}. Additionally, the $CP$-violating $\widetilde{b}_Z$ anomalous coupling may be induced through flavor changing neutral currents mediated by the $Z$ or $H$ bosons, as in the $HZ\gamma$ vertex \cite{Hernandez-Juarez:2024pty}.  

Since the  anomalous couplings can be taken as complex in general \cite{Hernandez-Juarez:2023dor,Hernandez-Juarez:2021xhy}, the form factors of \eqref{H11}-\eqref{H33} will be written as follows
\begin{equation}\label{reim}
h_i^H={\rm Re}\big[h_i^H\big]+i {\rm Im}\big[h_i^H\big].
\end{equation}

Before moving forward, it is important to note a few general limitations regarding the prescription used in Lagrangian \eqref{Lag} and the absorptive parts of the form factors. It is observed  that the $HZZ$ Lagrangian requires real $\hat{b}_Z$, $\hat{c}_Z$ and $\widetilde{b}_Z$ anomalous couplings to be Hermitian. This seems to contradict our assumption in Eq. \eqref{reim}. However, the Lagrangian \eqref{Lag} is only valid in the Born approximation as it exclusively describes the external particles in an effective approach \cite{Hagiwara:1986vm}. On the other hand, anomalous couplings are generated through quantum corrections by heavy or light particles at one-loop level or higher orders. The new operators that induce these loop effects are not accounted for in Lagrangian \eqref{Lag} and do not necessarily necessitate real anomalous couplings to form Hermitian amplitudes \cite{Hagiwara:2000tk}. For instance, in the SM, the contributions arising at the one-loop level develop an absorptive part as long as the magnitude of the off-shell Higgs boson four-momentum is above the $\|q\| = 2m_i$ threshold, where $m_i$ is the mass of the particles coupled to $H$. Therefore, in the most general scenario, the $\hat{b}_Z$, $\hat{c}_Z$, and $\tilde{b}_Z$ anomalous couplings are no longer real constants but complex functions of $q^2$. This general scenario is the one we will consider in our analysis.

Although the absorptive parts have been neglected in experimental analyses at the LHC,  their effects have been discussed in some theoretical works \cite{Godbole:2007cn,Hagiwara:2000tk,Rao:2020hel} and more recently  via a left-right asymmetry \cite{Hernandez-Juarez:2023dor}, which may be reconstructed in the final state of the $H^\ast\rightarrow ZZ\rightarrow \overline{\ell}_1\ell_1\overline{\ell}_2\ell_2$ decay. 
Polarization effects of the $Z$ gauge bosons on the Higgs boson decay $H\to ZZ$ have already been studied \cite{Maina:2020rgd,Maina:2021xpe}, though only non-anomalous $HZZ$ couplings and  one on-shell $Z$ gauge boson were considered. Similar  calculations were also  reported in \cite{Soni:1993jc,Buchalla:2013mpa,Berge:2015jra,He:2019kgh}. On the other hand, the scenario where both $Z$ gauge bosons are on-shell and the $HZZ$ coupling is purely real was studied in  Refs. \cite{Gao:2010qx,Bolognesi:2012mm,Phan:2022amy}. Finally, a restrictive scenario where not all the $HZZ$ anomalous couplings were taken as complex was considered in Ref. \cite{Godbole:2007cn}. 

In this work, we are interested in the most general scenario where all the  anomalous $HZZ$ couplings are complex  and calculate the  unpolarized and polarized off-shell $H^\ast\rightarrow ZZ\rightarrow \overline{\ell}_1\ell_1\overline{\ell}_2\ell_2$ decay width. We will then analyze the implications of the absorptive parts of the $HZZ$ anomalous couplings on some observables.

\subsection{Polarized Amplitudes}\label{polSec}

\begin{figure}[!hbt]
\begin{center}
\includegraphics[width=10cm]{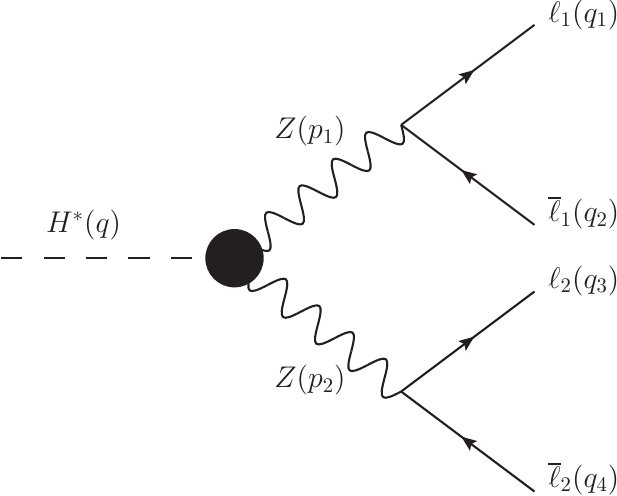}
\caption{ Nomenclature for the four-momenta of the particles involved in the Feynman diagram for the contribution of the $HZZ$ anomalous couplings to the Higgs boson decay  $H^\ast\rightarrow Z(p_1)Z(p_2)\rightarrow \overline{\ell}_1\ell_1\overline{\ell}_2\ell_2$. \label{KinFig}}
\end{center}
\end{figure}

Using the nomenclature of Fig. \ref{KinFig}, the  amplitude for the $H^\ast\rightarrow Z(p_1)Z(p_2)\rightarrow \overline{\ell}_1\ell_1\overline{\ell}_2\ell_2$ process can be written  as
\begin{align}
\label{amp1}
    \mathcal{M}_{H^\ast\rightarrow \overline{\ell}_1\ell_1\overline{\ell}_2\ell_2}=&\mathcal{M}^{\mu\nu}_{H^\ast\rightarrow Z(p_1)Z(p_2)} \frac{i\sum\epsilon_\eta(p_1,\lambda_1)\epsilon_\mu^{\ast}(p_1,\lambda_1)}{p_1^2-m_Z^2+i\Gamma_Z m_Z} \mathcal{M}^\eta_{Z(p_1)\rightarrow \overline{\ell}_1\ell_1} \frac{i\sum\epsilon_\kappa(p_2,\lambda_2)\epsilon^{\ast}_\nu(p_2,\lambda_2)}{p_2^2-m_Z^2+i\Gamma_Z m_Z}  \mathcal{M}^\kappa_{Z(p_2)\rightarrow \overline{\ell}_2\ell_2} ,
\end{align}
where $\Gamma_Z$ is the $Z$ gauge boson decay width, and, according to Lagrangian \eqref{Lag}, the amplitude for the process $H^\ast\rightarrow ZZ$ is given in terms of the complex $h_i^H$ form factors as follows  
\begin{align}
\mathcal{M}^{\mu\nu}_{H^\ast\rightarrow Z(p_1)Z(p_2)}=&i\Big(\frac{g }{c_W}\Big)m_Z\Big\{g^{\mu \nu}\Big({\rm Re}\big[h_1^H\big]+i {\rm Im}\big[h_1^H\big] \Big) +\frac{p_2^{\mu }{p_1}^{\nu }}{m_Z^2} \Big(  {\rm Re}\big[h_2^H\big]+i 
   {\rm Im}\big[h_2^H\big]\Big) \nonumber\\
 &+\frac{\epsilon^{\mu \nu \alpha \beta }p_{1\alpha} p_{2\beta} }{m_Z^2}\Big({\rm Re}\big[h_3^H\big]
  +i {\rm Im}\big[h_3^H\big] \Big)\Big\},
\end{align}
whereas the amplitudes for the $Z\rightarrow \overline{\ell}_1\ell_1$ and $Z\rightarrow \overline{\ell}_2\ell_2$ processes are
\begin{align}
    &  \mathcal{M}^\eta_{Z(p_1)\rightarrow \overline{\ell}_1\ell_1}=i\frac{g}{c_W}\overline{u}(q_2)\gamma^\eta \big( g_V-g_A \gamma^5\big)u(q_1),\\
    & \mathcal{M}^\kappa_{Z(p_2)\rightarrow \overline{\ell}_2\ell_2}=i\frac{g}{c_W}\overline{u}(q_4)\gamma^\kappa \big( g_V-g_A \gamma^5\big)u(q_3).
\end{align}

For the $Z$ gauge boson propagators of Eq. \eqref{amp1} we have used the completeness relation \cite{Hoppe:2023uux}
\begin{equation}\label{propagator}
-g^{\mu\nu}+\frac{k^\mu k^\nu}{m_Z^2}=\sum^3_{\lambda=1}\epsilon^\mu_{\lambda}(k)\epsilon^{\ast\nu}_{\lambda}(k),
\end{equation}
where we only consider 3 polarizations since both $Z$ gauge bosons are on-shell. 

In order to study the polarization effects in the final state, it is convenient to rewrite  Eq. \eqref{amp1} as a sum of the polarized amplitudes
\begin{align}
\label{amp2}
    &\mathcal{M}_{H^\ast\rightarrow \overline{\ell}_1\ell_1\overline{\ell}_2\ell_2}(\lambda_1,\lambda_2)=\sum_{\lambda_1}\sum_{\lambda_2}\frac{\mathcal{M}_{H^\ast\rightarrow Z(p_1)Z(p_2)}(\lambda_1,\lambda_2) \mathcal{M}_{Z(p_1)\rightarrow \overline{\ell}_1\ell_1}(\lambda_1)\mathcal{M}_{Z(p_2)\rightarrow \overline{\ell}_2\ell_2}(\lambda_2)}{\big(p_1^2-m_Z^2+i\Gamma m_Z\big)\big(p_2^2-m_Z^2+i\Gamma m_Z\big)}
\end{align}
where
\begin{align}
\mathcal{M}_{H^\ast\rightarrow Z(p_1)Z(p_2)}(\lambda_1,\lambda_2)&=\mathcal{M}_{H^\ast\rightarrow Z(p_1)Z(p_2)}^{\mu\nu}\epsilon^\ast_\mu(p_1,\lambda_1)\epsilon_\nu^\ast(p_2,\lambda_2),\\
  \mathcal{M}_{Z(p_i)\rightarrow \overline{\ell}_i\ell_i}(\lambda_i)&=\mathcal{M}_{Z(p_i)\rightarrow \overline{\ell}_i\ell_i}^\mu\epsilon_\mu(p_i,\lambda_i).
\end{align}

Since the $\mathcal{M}_{H^\ast\rightarrow Z(p_1)Z(p_2)}(\lambda_1,\lambda_2)$ polarized amplitude is  non-vanishing for $\lambda_1=\lambda_2$ only \cite{Hernandez-Juarez:2023dor,Bolognesi:2012mm}, the square amplitude can be written as 
\begin{align}
\label{ampT1}
        \mathcal{M}^2_{H^\ast\rightarrow \overline{\ell}_1\ell_1\overline{\ell}_2\ell_2}=   &\frac{\sum_\lambda \mathcal{M}^2_{H^\ast\rightarrow Z(p_1)Z(p_2)}(\lambda,\lambda) \mathcal{M}^2_{Z(p_1)\rightarrow \overline{\ell}_1\ell_1}(\lambda)\mathcal{M}^2_{Z(p_2)\rightarrow \overline{\ell}_2\ell_2}(\lambda)+\mathcal{M}^2_{\rm int}
        }{\left((p_1^2-m_Z^2)^2+\Gamma^2 m_Z^2\right)\left((p_2^2-m_Z^2)^2+\Gamma^2 m_Z^2\right)}.
\end{align}
 where the interference term $\mathcal{M}^2_{\rm int}$ is given by
\begin{align}\label{suminter}
\mathcal{M}^2_{\rm int}=&\sum_{\lambda_1} \sum_{\lambda_2\neq \lambda_1 }\mathcal{M}_{H^\ast\rightarrow Z(p_1)Z(p_2)}(\lambda_1,\lambda_1)         \mathcal{M}^\dagger_{H^\ast\rightarrow Z(p_1)Z(p_2)}(\lambda_2,\lambda_2)\mathcal{M}_{Z(p_1)\rightarrow \overline{\ell}_1\ell_1}(\lambda_1)\mathcal{M}^\dagger_{Z(p_1)\rightarrow \overline{\ell}_1\ell_1}(\lambda_2)\nonumber\\
&\times\mathcal{M}_{Z(p_2)\rightarrow \overline{\ell}_2\ell_2}(\lambda_1)\mathcal{M}^\dagger_{Z(p_2)\rightarrow \overline{\ell}_2\ell_2}(\lambda_2).
\end{align}
From the first term of the right-hand side of Eq. \eqref{ampT1} we can extract the contributions of polarized $Z$ gauge bosons with polarizations $\lambda=L$, $R$, and $0$, which allows one to study the polarized  decay width and other observables of interest.
As described below, to obtain the $H^\ast\rightarrow \overline{\ell}_1\ell_1\overline{\ell}_2\ell_2$ decay width we will follow the approach  of Ref. \cite{Buchalla:2013mpa} for the calculation of the $H\rightarrow ZZ^\ast\rightarrow\overline{\ell}_1\ell_1\overline{\ell}_2\ell_2$ decay. A similar procedure was followed in Ref. \cite{Cappiello:2011qc,Gevorkyan:2014waa} for the decay $K^\pm\rightarrow \pi^\pm \pi^0 e^+ e^-$, whereas an alternative method was implemented in Ref. \cite{Berge:2015jra} for the  $H\rightarrow ZZ^\ast\rightarrow\ell^+\ell^-\tau^+\tau^-$ and $H\rightarrow WW^\ast\rightarrow\ell^-\overline{\nu}_\ell \tau^+\nu_\tau$ decays using transformation properties of the helicity amplitudes under the rotation group.

\begin{figure}[!hbt]
\begin{center}
{\includegraphics[width=12cm]{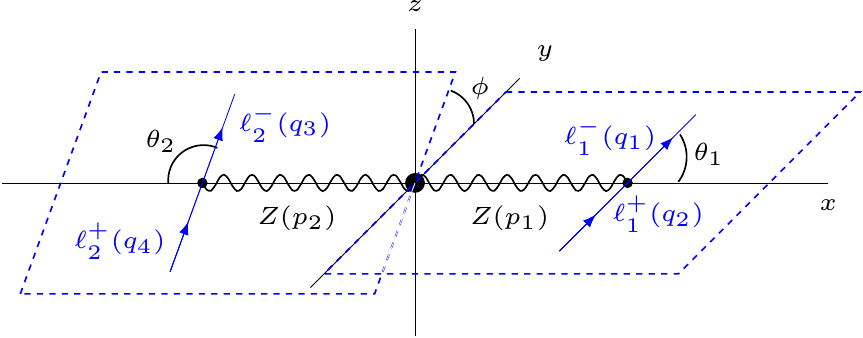}}
\caption{Nomenclature for the reference systems and angles used in the calculation of the $H\to ZZ\to \bar{\ell}_1\ell_1\bar{\ell}_2\ell_2$ decay, as described in Appendix \ref{kinematics}.\label{plane}} 
\end{center}
\end{figure}

The  phase space for the $H^\ast\rightarrow \overline{\ell}_1\ell_1\overline{\ell}_2\ell_2$ process, along with the kinematics, is presented in Appendix \ref{kinematics}.
The relevant angular variables on which the square $\mathcal{M}^2_{H^\ast\rightarrow \overline{\ell}_1\ell_1\overline{\ell}_2\ell_2}$ amplitude depends are $\theta_1$, $\theta_2$ and $\phi$, which are described in Fig \ref{plane}. We note that  the square polarized amplitudes  $\mathcal{M}^2_{H^\ast\to ZZ}(\lambda,\lambda)$, $\mathcal{M}^2_{Z(p_1)\to \bar{\ell}_1\ell_1}(\lambda)$ and $\mathcal{M}^2_{Z(p_2)\to \bar{\ell}_2\ell_2}(\lambda)$ can be computed in the corresponding $H^\ast$, $Z(p_1)$ and $Z(p_2)$ rest frames since each one depends only on  scalar products of four-momenta given in the same inertial frame. Even more, these square amplitudes are  Lorentz invariant and  it is not necessary to apply a boost to bring them into a common reference system. 
On the other hand, the interference term $\mathcal{M}^2_{\rm int}$ is given in terms of partial amplitudes evaluated in distinct reference frames, so a boost from the $Z(p_i)\rightarrow\overline{\ell}_i\ell_i$ reference frames into the Higgs boson rest frame will be required to perform the sum of Eq. \eqref{suminter}. The same is true for the calculation of the unpolarized case via standard techniques as scalar products of  four-momenta given in distinct reference frames are also involved. 

We present below the analytical expressions for the polarized square amplitudes  and  the interference term of Eq. \eqref{ampT1}, which were calculated using the FeynCalc package \cite{Mertig:1990an,Shtabovenko:2016sxi,Shtabovenko:2020gxv}. The relation $p_i^2=m_Z^2$ ($i=1$, 2) will be considered  as the energy region where both $Z$ gauge bosons are on-shell is of interest for our analysis.

\subsubsection{Polarized $H^\ast\rightarrow Z(p_1)Z(p_2)$ square amplitudes}

The polarized square  amplitudes $\mathcal{M}^2_{H^\ast}(\lambda,\lambda)$ given in terms of the real and absorptive parts of the complex anomalous couplings $h_i^H$ of Lagrangian \eqref{Lag} were presented by some of us in a previous work \cite{Hernandez-Juarez:2023dor}. We consider a reference frame in which the motion of the $Z(p_1)$ gauge boson is along the positive direction of the $x$ axis and the off-shell Higgs boson is at rest. For further details see Appendix \ref{kinematics}. In the energy region where both $Z$ gauge bosons are on-shell, the transversally  polarized  $H^\ast\rightarrow ZZ$ amplitudes read as follows

\begin{align}
\label{MLL}
 \mathcal{M}^2_{H^\ast\rightarrow Z(p_1)Z(p_2)}(L,L)&= \left(\frac{g}{2c_W m_Z}\right)^2\left(4  m_Z^2  Q\sqrt{Q^2-4  m_Z^2}\, {\rm Im}\left(h_3^H {h_1^H}^\dagger\right)+Q^2 \left(Q^2-4
   m_Z^2\right) \big|h_3^H\big|^2+4 m_Z^4\big|h_1^H\big|^2\right),
\end{align}
and
\begin{align}
\label{MRR}
 \mathcal{M}^2_{H^\ast\rightarrow Z(p_1)Z(p_2)}(R,R)&=\left(\frac{g}{2c_W m_Z}\right)^2\left(-4  m_Z^2  Q\sqrt{Q^2-4 m_Z^2}\, {\rm Im}\left(h_3^H {h_1^H}^\dagger\right)+Q^2 \left(Q^2-4
   m_Z^2\right) \big|h_3^H\big|^2+4 m_Z^4
   \big|h_1^H\big|^2\right),
   \end{align}
whereas for the longitudinal polarization we obtain
\begin{align}\label{M00}
 \mathcal{M}^2_{H^\ast\rightarrow Z(p_1)Z(p_2)}(0,0)&=\left(\frac{g}{4c_W m^3_Z}\right)^2\Bigg(4 Q^6 m_Z^2 \left({\rm Re}\left(h_1^H {h_2^H}^\dagger\right)-2 \big|h_2^H\big|^2\right)+4 Q^4 m_Z^4 \left(\big|h_1^H\big|^2+4
   \big|h_2^H\big|^2-6{\rm Re}\left(h_1^H {h_2^H}^\dagger\right)\right)\nonumber\\
   &-16 Q^2 m_Z^6
   \left(\big|h_1^H\big|^2-2{\rm Re}\left(h_1^H {h_2^H}^\dagger\right)\right)+16 m_Z^8
   \big|h_1^H\big|^2+Q^8 \big|h_2^H\big|^2\Bigg),
   \end{align}  
where we have introduced  the four-lepton invariant mass $Q^2=q^2$, instead of the four-momentum of the off-shell Higgs boson. Moreover, we use ${\rm Re}\left(h_i^H {h_j^H}^\dagger\right)={\rm Re}\big[h_i\big]{\rm Re}\big[h_j\big]+{\rm Im}\big[h_i\big]{\rm Im}\big[h_j\big]$ and ${\rm Im}\left(h_i^H {h_j^H}^\dagger\right)={\rm Im}\big[h_i\big]{\rm Re}\big[h_j\big]-{\rm Re}\big[h_i\big]{\rm Im}\big[h_j\big]$. Our results reproduce  those reported in Ref. \cite{Bolognesi:2012mm} for the scenario with real anomalous couplings. 

Note that there is no dependence on the angular variables. Also, a non-vanishing $h_3^H$ form factor may give rise to  a left-right asymmetry in the four-lepton final state, which vanishes up to the one-loop level in the SM.

\subsubsection{$Z(p_i)\rightarrow 2\ell_i$ square amplitudes}
   
As already mentioned, the square amplitudes $\mathcal{M}^2_{Z(p_1)\rightarrow \overline{\ell}_1\ell_1}(\lambda)$ and $\mathcal{M}^2_{Z(p_2)\rightarrow \overline{\ell}_2\ell_2}(\lambda)$  can be worked out in the rest frame of the $Z(p_i)$ gauge boson. According to the kinematics presented in Appendix \ref{kinematics}, the polarization vectors of the $Z(p_1)$ ($Z(p_2)$) gauge boson are oriented along the positive (negative) direction of the $x$ axis, thus the transversally and longitudinally polarized amplitudes  are given as 
\begin{align}
\label{MZLR1}
 \mathcal{M}^2_{Z(p_i)\rightarrow \overline{\ell}_i\ell_i}(R/L)&= \left(\frac{g}{c_W} \right)^2 \left( s^2_{\theta_i} \left(4 m_i^2-m_Z^2\right) \left(g_A^2+g_V^2\right)\pm4 g_A
   g_V c_{\theta_i} m_Z\sqrt{m_Z^2-4 m_i^2 }+2 m_Z^2
   \left(g_A^2+g_V^2\right)-8 m_i^2 g_A^2 \right)\nonumber\\&\simeq
 \left( \frac{g}{c_W}\right)^2m_Z^2 \left(  \left(g_A^2+g_V^2\right)
   \left(1+ c^2_{\theta _i}\right)\pm4
   g_A g_V c_{\theta
   _i} \right),   
\end{align}
and
\begin{align}
\label{MZ00}
 \mathcal{M}^2_{Z(p_i)\rightarrow \overline{\ell}_i\ell_i}(0)&= 2\left(\frac{g}{c_W} \right)^2 \left( c^2_{\theta_i} \left(4 m_i^2-m_Z^2\right)
   \left(g_A^2+g_V^2\right)+m_Z^2 \left(g_A^2+g_V^2\right)-4 m_i^2 g_A^2\right)\nonumber\\
   &\simeq 2\left( \frac{g}{ c_W}\right)^2  s^2_{\theta _i} m_Z^2 \left(g_A^2+g_V^2\right),
\end{align}
where $i=1,2$ and we have introduced the short-hand notation $c_\eta\equiv \cos\eta$ and $s_\eta\equiv \sin\eta$. The approximate results were obtained in the massless lepton limit, which is used in the 
phase space defined in Appendix \ref{kinematics}.

To cross-check our calculation, we have boosted Eqs. \eqref{MZLR1} and \eqref{MZ00}   into the Higgs boson rest frame via the Lorentz transformation given in Appendix \ref{kinematics},  which leaves them invariant as expected. 
We also note that the above square amplitudes are independent of the $\phi$ angle and the four-lepton invariant mass $Q$. Also, although the $g_Ag_V$ term in Eq. \eqref{MZLR1} seems to give rise to a  left-right asymmetry, it vanishes after  integration over the $\theta_i$ angles. Therefore, the only  observable  effects of the left- and right-handed polarized  $Z(p_i)\rightarrow\overline{\ell}_i\ell_i$ amplitudes would appear via either angular distributions or a forward-backward asymmetry.   
   \subsubsection{Interference term}

For the purpose of our work, it is enough to consider the massless lepton approximation, which yields

\begin{align}
\label{inter}
\mathcal{M}^2_{\rm int}&\equiv \sum_{\lambda_1} \sum_{\lambda_2\neq \lambda_1 } \mathcal{M}^2_{H^\ast Z(p_1)Z(p_2)}(\lambda_1,\lambda_2) 
\nonumber\\&= 
 \frac{s_{\theta_1} s_{\theta_2}}{2m_Z^2}\Big(
 s_{\theta_1} s_{\theta_2} m_Z^2\left(g_A^2+g_V^2\right){}^2 f_1\left(Q^2\right)
 +\big(2 g_A g_V+c_{\theta _1}
   \left(g_A^2+g_V^2\right)\big)
   \big(2 g_A g_V+c_{\theta_2}
   \left(g_A^2+g_V^2\right)\big) f_2\left(Q^2\right)\nonumber\\&
      +\big(2 g_A g_V-c_{\theta _1}
   \left(g_A^2+g_V^2\right)\big)
   \big(2 g_A g_V-c_{\theta_2}
   \left(g_A^2+g_V^2\right)\big) f_3\left(Q^2\right)\Big),
\end{align}   
where
\begin{align} 
f_1\left(Q^2\right)&= 4  \sqrt{Q^2\left(Q^2-4 m_Z^2\right)}\,
   m_Z^2
   {\rm Re}\left(h_1^H {h_3^H}^\dagger\right)s_{2 \phi}+\left(4m_Z^4\big|h_1^H\big|^2 -Q^2
   \left(Q^2-4
   m_Z^2\right)\big|h_3^H\big|^2\right) c_{2 \phi },
\end{align}   
   
\begin{align}   
f_2\left(Q^2\right)&=c_{\phi}
   \Bigg(
   \left(Q^2\left( Q^2-4 m_Z^2\right) \right){}^{3/2}{\rm Im}\left(h_3^H {h_2^H}^\dagger\right)
   \nonumber\\
   &+2 m_Z^2
   \left(
   \sqrt{ Q^2\left( Q^2-4 m_Z^2 \right)} \left(Q^2-2
   m_Z^2\right){\rm Im}\left(h_3^H{h_1^H}^\dagger\right)+ \left(4 
   m_Z^2-Q^2\right)Q^2{\rm Re}\left(h_1^H{h_2^H}^\dagger\right)\right)\Bigg)\nonumber\\
   &- s_{\phi}
   \Bigg(
   \left( Q^2\left( Q^2-4 m_Z^2\right)\right){}^{3/2}{\rm Re}\left(h_2^H{h_3^H}^\dagger\right)
   \nonumber\\
   &+2m_Z^2
   \left(
   \sqrt{Q^2-4 m_Z^2} \left(Q^2-2
   m_Z^2\right){\rm Re}\left(h_1^H{h_3^H}^\dagger\right)+ \left(Q^2-4 
   m_Z^2\right)Q^2{\rm Im}\left(h_1^H{h_2^H}^\dagger\right)\right)\Bigg)\nonumber\\
   &+4c_{\phi}m_Z^4\left(2 m_Z^2- Q^2\right)\big|h_1^H\big|^2,   
\end{align}   
and   
\begin{align}
f_3\left(Q^2\right)&= c_{\phi }
   \Bigg(-
   \left(Q^2\left(Q^2-4 m_Z^2\right)\right){}^{3/2}{\rm Im}\left(h_3^H {h_2^H}^\dagger\right)
   \nonumber\\
   &+2 m_Z^2
   \left(
   \sqrt{Q^2\left(Q^2-4 m_Z^2\right)} \left(2
   m_Z^2-Q^2\right){\rm Im}\left(h_3^H{h_1^H}^\dagger\right)+ \left(4 
   m_Z^2-Q^2\right)Q^2{\rm Re}\left(h_1^H{h_2^H}^\dagger\right)\right)\Bigg)\nonumber\\
   &-s_{\phi}
   \Bigg(   
     \left(Q^2\left(Q^2-4 m_Z^2\right)\right){}^{3/2}{\rm Re}\left(h_2^H{h_3^H}^\dagger\right)\nonumber\\
   &+2  m_Z^2 \left(  
   \sqrt{Q^2-4 m_Z^2} \left(Q^2-2
   m_Z^2\right){\rm Re}\left(h_1^H{h_3^H}^\dagger\right)+ \left(4
   m_Z^2-Q^2\right)Q^2{\rm Im}\left(h_1^H{h_2^H}^\dagger\right)\right)\Bigg)
   \nonumber\\
   &+4c_{\phi }m_Z^4
   \left(2 m_Z^2-4 Q^2\right)\big|h_1^H\big|^2.
\end{align}
Although the terms dependent  on the $\phi$ angle vanish after integration in the phase space, the study of observables related to  $\phi$ can lead to interesting results.

\subsection{$H^\ast\rightarrow Z(p_1)Z(p_2)\rightarrow\overline{\ell}_1\ell_1\overline{\ell}_2\ell_2$ decay width}

Following the discussion of Appendix \ref{kinematics}, the differential $H^\ast\rightarrow Z(p_1)Z(p_2)\rightarrow\overline{\ell}_1\ell_1\overline{\ell}_2\ell_2$  decay width can  be written as
\begin{equation}
\label{difEq1}
\frac{d\Gamma_{H^\ast\rightarrow \overline{\ell}_1\ell_1\overline{\ell}_2\ell_2}}{dp^2_1dp^2_2d\cos{\theta_1}d\cos{\theta_2}d\phi}=\frac{\sqrt{Q^2-4 m_Z^2}}{ 512(2\pi)^6 Q^2 }         \mathcal{M}^2_{H^\ast\rightarrow \overline{\ell}_1\ell_1\overline{\ell}_2\ell_2}.
\end{equation}
 
Since we consider the scenario where both $Z$ gauge bosons are on-shell, we use the narrow-width approximation
for their  Breit-Wigner propagators
\begin{equation}
\label{NarrowApprox}
\lim_{m_Z \Gamma_Z\rightarrow 0}\frac{1}{(p_i^2-m_Z^2)^2+(m_Z\Gamma_Z)^2}=\delta(p_i^2-m_Z^2)\frac{\pi}{m_Z\Gamma_Z},\quad (i=1,2),
\end{equation}
which allows one to integrate over $p_1$ and $p_2$. Introducing the polarized and interference amplitudes the  differential decay width can be written as
\begin{align}
\label{difEq2}
\frac{d\Gamma_{H^\ast\rightarrow \overline{\ell}_1\ell_1\overline{\ell}_2\ell_2}}{d\cos{\theta_1}d\cos{\theta_2}d\phi}=&\frac{\sqrt{Q^2-4  m_Z^2}}{(32\sqrt{2})^2 (2\pi)^4 Q^2 (m_Z\Gamma_Z)^2}         \Big(\sum_\lambda \mathcal{M}^2_{H^\ast\rightarrow Z(p_1)Z(p_2)}(\lambda,\lambda) \mathcal{M}^2_{Z(p_1)\rightarrow \overline{\ell}_1\ell_1}(\lambda)\mathcal{M}^2_{Z(p_2)\rightarrow \overline{\ell}_2\ell_2}(\lambda)\nonumber\\
        &+ \mathcal{M}^2_{\rm int}\Big).
\end{align}
All our results presented below will be obtained from the above equation along with Eqs. \eqref{MLL}--\eqref{MZ00}, whereas the interference term  \eqref{inter}  will be integrated out as we are mainly interested in the polarization effects.

Below we present the  $H^\ast\rightarrow \overline{\ell}_1\ell_1\overline{\ell}_2\ell_2$ decay width  in the scenario with polarized  $Z$ gauge bosons, though for completeness we also consider the case of unpolarized $Z$ gauge bosons, which can be obtained via standard calculation techniques and serve  to cross-check our results for polarized $Z$ gauge bosons obtained from Eq. \eqref{difEq2}.

\subsubsection{Polarized $H^\ast\rightarrow \overline{\ell}_1\ell_1\overline{\ell}_2\ell_2$ decay width}

To obtain the polarized $H^\ast\rightarrow \overline{\ell}_1\ell_1\overline{\ell}_2\ell_2$ decay width we  integrate  Eq. \eqref{difEq2} over the angular variables in the region given in Appendix \ref{kinematics}, which after some rearrangement yields

\begin{align}
\label{TotalWidthPol}
\Gamma^\lambda_{H^\ast\rightarrow \overline{\ell}_1\ell_1\overline{\ell}_2\ell_2}=&\frac{g^6
   \left(g_A^2+g_V^2\right){}^2
   \sqrt{Q^2-4 m_Z^2}}{
   (48\sqrt{2})^2 (2\pi)^3 Q^2 c_W^6 m_Z^4
   \Gamma _Z^2}F^\lambda\left(Q^2\right),
\end{align}   
with
\begin{align}
F^\lambda\left(Q^2\right)&=
\left(
   \left(Q^2-2 m_Z^2\right){}^2 \left(4 m_Z^4\big|h_1^H\big|^2+Q^4
  \big|h_2^H\big|^2\right)
   +4 Q^2 m_Z^2 \left(Q^4-6
   Q^2 m_Z^2+8 m_Z^4\right){\rm Re}\left(h_1^H{h_2^H}^\dagger\right)\right)f_0\nonumber\\&+    
    4
   m_Z^4 \left(4 m_Z^2
   \sqrt{Q^2\left(Q^2-4 m_Z^2\right)}{\rm Im}\left(h_3^H{h_1^H}^\dagger\right)+Q^2
   \left(Q^2-4 m_Z^2\right)\big|h_3^H\big|^2+
   4 m_Z^4\big|h_1^H\big|^2\right)f_L
   \nonumber\\ 
   &+ 4m_Z^2\left(-4 m_Z^2
   \sqrt{Q^2\left(Q^2-4 m_Z^2\right)}{\rm Im}\left(h_3^H{h_1^H}^\dagger\right)
   +Q^2
   \left(Q^2-4 m_Z^2\right)
   \big|h_3^H\big|^2+4 m_Z^4\big|h_1^H\big|^2\right)f_R,
    \end{align}
where  the $f_L$, $f_R$ and $f_0$ coefficients must be set to $0$ or $1$ according to the polarization  $\lambda$ of the  $Z$ gauge bosons. For instance,  $f_L=1$ and $f_R=f_0=0$ correspond to $\lambda=L$.
As already pointed out,  the interference term in Eq.  \eqref{difEq2} vanishes after  $\phi$ integration. Thus, it is possible to discriminate between the distinct contributions of polarized $Z$ gauge bosons, which are functions of the four-lepton invariant mass only.   

The polarized $H^\ast\rightarrow \overline{\ell}_1\ell_1\overline{\ell}_2\ell_2$ decay width can also be written in terms of the $Z(p_i)\rightarrow\overline{\ell}_i\ell_i$ branching fractions as follows
   \begin{align}
\label{TotWpol}
\Gamma^\lambda_{H^\ast\rightarrow \overline{\ell}_1\ell_1\overline{\ell}_2\ell_2}&=2\ \Gamma^\lambda_{H^\ast\rightarrow ZZ}\left(Q^2\right) \text{Br}(Z\rightarrow \ell_1\overline{\ell}_1)\text{Br}(Z\rightarrow \ell_2\overline{\ell}_2)=0.00226418\times \Gamma^\lambda_{H^\ast\rightarrow Z_\lambda Z_\lambda}\left(Q^2\right),
\end{align}
where an extra factor of 2 is included in Eq. \eqref{TotWpol},  which stems from the fact that the $Z$ bosons are not in the final state, and hence we have not considered the statical factor of 1/2 that corrects the double-counting of identical particles in the $H^\ast\rightarrow ZZ$ amplitude. This factor has been  overlooked in the past \cite{Phan:2022amy}. In. Eq. \eqref{TotWpol}, the polarized $H^\ast\rightarrow  Z_\lambda Z_\lambda$ decay width is given by
\begin{align}\label{HZZpolG}
\Gamma^\lambda_{H^\ast\rightarrow Z_\lambda Z_\lambda}&=\frac{g^2
   \sqrt{Q^2-4 m_Z^2}}{
   128 \pi Q^2 c_W^2 m_Z^6
  }F^\lambda\left(Q^2\right).
   \end{align}
and we have used
\begin{align}
\label{ZlepW}
\text{Br}(Z\rightarrow \ell_i\overline{\ell}_i)&\simeq\frac{g^2 m_Z }{12 \pi  c_W^2\Gamma_Z}\left(g_A^2+g_V^2\right)\simeq 0.0336466,
\end{align}
in the massless lepton approximation  \cite{ParticleDataGroup:2022pth}.

From \eqref{TotWpol} it is clear that the four-lepton final state is sensitive to the polarization effects on the $HZZ$ vertex. Therefore, the left-right asymmetry  in  $Z$ gauge boson pair production \citep{Hernandez-Juarez:2023dor} has also effects on the decay of an off-shell Higgs boson into  four-leptons. Such an asymmetry is a consequence of $CP$ violation and complex anomalous couplings. Hence, new physics effects may be detected via  transversally polarized $Z$ gauge bosons \cite{Lee:2018fxj}. Moreover, they would provide an excellent probe of quantum field theory, which predicts that the $h_i^V$ ($i$=1, 2, 3) form factors are complex \cite{Hernandez-Juarez:2021xhy}.  On the other hand, longitudinally polarized $Z$ gauge bosons are not useful to detect   $CP$-violating effects, though it is still possible to detect new-physics contributions through the $\hat{c}_Z$ anomalous coupling, which enters into the  definition of the $h_1^H$  form factor. 

\subsubsection{Unpolarized  $H^\ast\rightarrow \overline{\ell}_1\ell_1\overline{\ell}_2\ell_2$ decay width}
For the sake of completeness, we also consider unpolarized $Z$ gauge bosons.
From Eq. \eqref{TotalWidthPol}, we can obtain the unpolarized $H^\ast\rightarrow \overline{\ell}_1\ell_1\overline{\ell}_2\ell_2$ decay width by setting $f_L=f_R=f_0=1$, which yields
\begin{align}
\label{TotalWidthUnPol}
\Gamma_{H^\ast\rightarrow \overline{\ell}_1\ell_1\overline{\ell}_2\ell_2}\left(Q^2\right)&=    \frac{g^6
   \left(g_A^2+g_V^2\right){}^2
   \sqrt{Q^2-4 m_Z^2} }{(48\sqrt{2})^2 (2\pi)^3 Q^2 c_W^6 m_Z^4 \Gamma
   _Z^2}G\left(Q^2\right),
\end{align}
where   
\begin{align}
\label{Gfunc}
G^\lambda\left(Q^2\right)&=   Q^2
   \left(Q^2-4 m_Z^2\right)
   \left(Q^2 \left(Q^2-4
   m_Z^2\right)
    \big|h_2^H\big|^2+8 m_Z^4
    \big|h_3^H\big|^2\right)
   +4 m_Z^4
   \left(Q^4-4 Q^2 m_Z^2+12
   m_Z^4\right) \big|h_1^H\big|^2\nonumber\\&+4 Q^2 m_Z^2
    \left(Q^4-6
   Q^2 m_Z^2+8 m_Z^4\right)
   {\rm Re}\left(h_1^H {h_2^H}^\dagger\right).
\end{align}
To cross-check  this result, we have inserted the left-hand side of Eq. \eqref{propagator} into Eq. \eqref{amp1} and used standard calculation techniques. In this case, the boosts defined in Appendix \ref{kinematics} are necessary as the square amplitude depends on  scalar products of  four-momentum given in their respective $Z(p_i)$ gauge boson rest frame. 

From Eq. \eqref{TotalWidthUnPol} we can observe that  the contribution of the $CP$-violating form factor $h_3^H$ to the four-lepton final state is suppressed as compared to the $CP$-conserving contributions, which  are proportional to higher powers of the four-lepton invariant mass $Q$. Thus, the $H^\ast\rightarrow \overline{\ell}_1\ell_1\overline{\ell}_2\ell_2$ unpolarized decay width is not really sensitive to $CP$-violating  effects. We also note that one can extract from Eq. \eqref{TotalWidthUnPol} the  tree-level SM contribution, which  is obtained for $h_2^H=h_3^H=0$ and $h_1^H=1$. As far as the one-loop level SM contribution is concerned, it is induced by setting $h_3^H=0$, whereas the remaining form factors are given in terms  of $\hat{b}_Z$ only since $\hat{c}_Z$ vanishes at this order. The one-loop level contribution to $\hat b_Z$ was reported in Ref. \cite{Hernandez-Juarez:2023dor}.

Eq. \eqref{TotalWidthUnPol}  can be written in the following short form  
   \begin{align}
\label{TotW}
\Gamma_{H^\ast\rightarrow \overline{\ell}_1\ell_1\overline{\ell}_2\ell_2}&=2\ \Gamma_{H^\ast\rightarrow ZZ}\left(Q^2\right) \text{Br}(Z\rightarrow \ell_1\overline{\ell}_1)\text{Br}(Z\rightarrow \ell_2\overline{\ell}_2)=0.00226418\times \Gamma_{H^\ast\rightarrow ZZ}\left(Q^2\right) ,
\end{align}
where the unpolarized $H^\ast\rightarrow ZZ$  decay width  for complex anomalous couplings is given by
\begin{align}
\label{WidthHZZ}
    \Gamma_{H^\ast\rightarrow ZZ}\left(Q^2\right)=& \frac{g^2 \sqrt{Q^2-4  m_Z^2}}{512 \pi 
   Q^2 c_W^2 m_Z^6}G\left(Q^2\right). 
\end{align}
The inclusion of a factor of $2$  in Eq. \ref{TotW} has already been explained when obtaining the polarized decay width. The unpolarized $H^\ast\rightarrow ZZ \rightarrow  \overline{\ell}_1\ell_1\overline{\ell}_2\ell_2$ process has been studied long ago. Thus, we can find both theoretical and numerical results in the literature \cite{Bredenstein:2006rh,Bredenstein:2006nk,Bredenstein:2006ha,Altenkamp:2017ldc,Altenkamp:2017kxk,Altenkamp:2018bcs}.

\section{Numerical analysis}
\label{numanal}
We now turn to the study of the polarized $H^\ast\rightarrow \overline{\ell}_1\ell_1\overline{\ell}_2\ell_2$ decay width and a few observables sensitive to new physics effects. We first present a cross-check  of our numerical evaluation method, which relies on the analytical results presented in Sec. \ref{framework} along with the kinematics and four-body phase space discussed in Appendix \ref{kinematics}.

\subsection{Consistency of our numerical evaluation method\label{SecVal}}

Using our analytical results, we have implemented a Mathematica code to perform a numerical evaluation, which is available in our GitLab repository \cite{urlcode}. For an alternative numerical evaluation, we have implemented the Lagrangian of Eq. \eqref{Lag} into \texttt{MadGraph5\_aMC@NLO} \cite{Alwall:2014hca} via the UFO format \cite{Degrande:2011ua} using the FeynRules package \cite{Christensen:2008py}. 
We then compare the numerical results  obtained via these two independent methods. The $e^-e^+\mu^-\mu^+$ final state is considered in our analysis, which, however, is indistinguishable from other $\bar{\ell}_1 \ell_1 \bar{\ell}_2 \ell_2$  final states as the massless lepton approximation is used.  

\begin{figure}[!hbt]
\begin{center}
{\includegraphics[width=9cm]{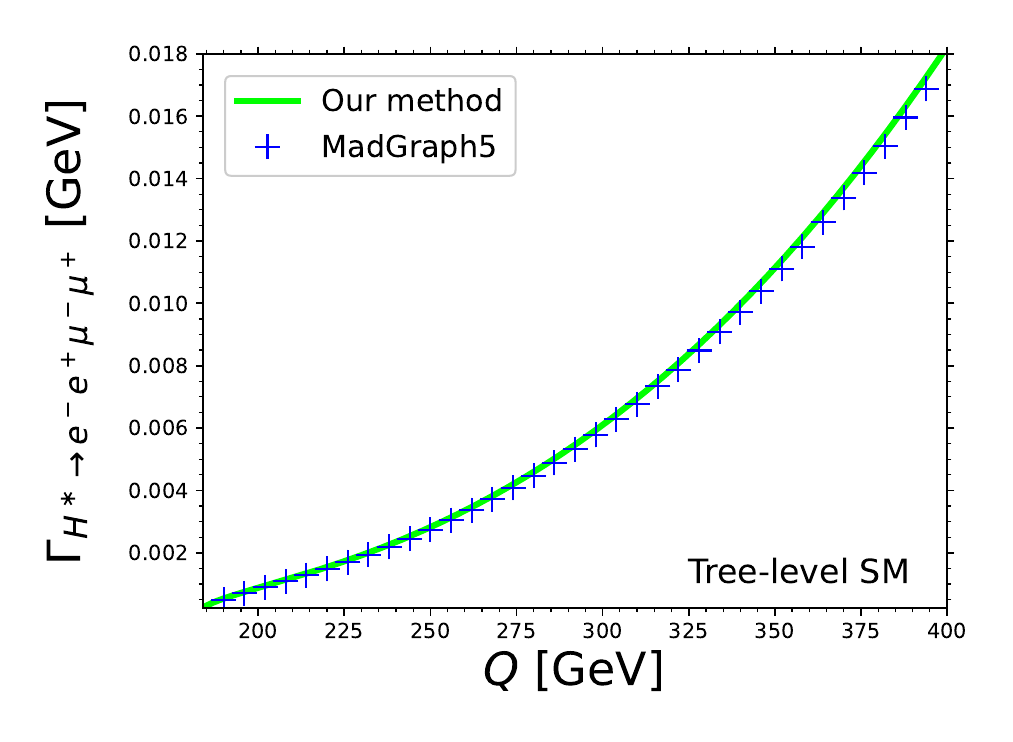}}
{\includegraphics[width=8.5cm]{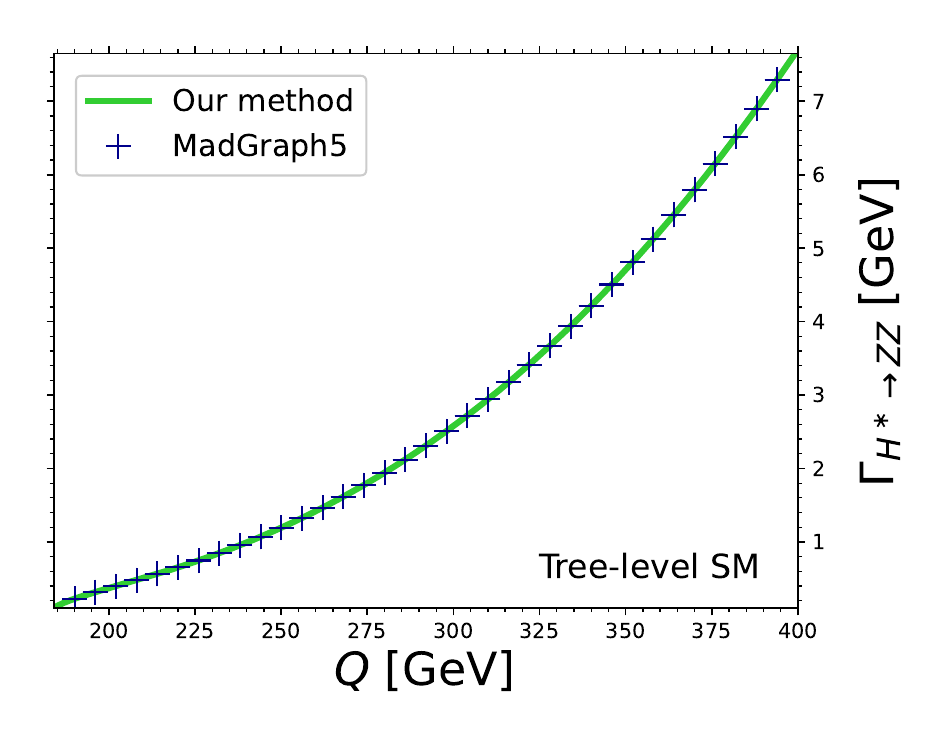}}
\caption{Tree-level SM contribution to the unpolarized $H^\ast\rightarrow ZZ \rightarrow e^-e^+\mu^-\mu^+$ and  $H^\ast\rightarrow ZZ $ decay widths as functions of the four-lepton invariant mass $Q$ obtained by both our own evaluation method and  \texttt{MadGraph5\_aMC@NLO}. \label{unpdecaywidths}} 
\end{center}
\end{figure}

Since the anomalous $HZZ$ couplings have negligible effects on the unpolarized $H^*\to \bar{\ell}_1 \ell_1 \bar{\ell}_2 \ell_2$   decay width \cite{Hernandez-Juarez:2023dor}, we set $h_1^H=1$ and $h^H_2=h_3^H=0$ and calculate the tree-level SM contribution to the unpolarized $H^\ast\rightarrow ZZ \rightarrow e^-e^+\mu^-\mu^+$ and $H^\ast\rightarrow ZZ$ decays. The results  obtained by our own numerical evaluation method and  \texttt{MadGraph5\_aMC@NLO} are shown in  Fig. \ref{unpdecaywidths} as  functions of the four-lepton invariant mass. An excellent agreement between both results is observed, with a slight variation of the order  of 2 \% of the  decay widths.  We  have also verified that the $H^\ast\rightarrow ZZ \rightarrow e^-e^+\mu^-\mu^+$ decay width obtained via Eq. \eqref{TotW} agree with that obtained from Eq. \eqref{TotalWidthUnPol} and shown in the left plot of Fig. \ref{unpdecaywidths}, which in turn  shows the consistency of the extra  factor of 2 inserted into Eq. \eqref{TotW}. Finally, our numerical results agree with those reported in Ref. \cite{Bredenstein:2006rh}.

\begin{figure}[!hbt]
\begin{center}
{\includegraphics[width=9cm]{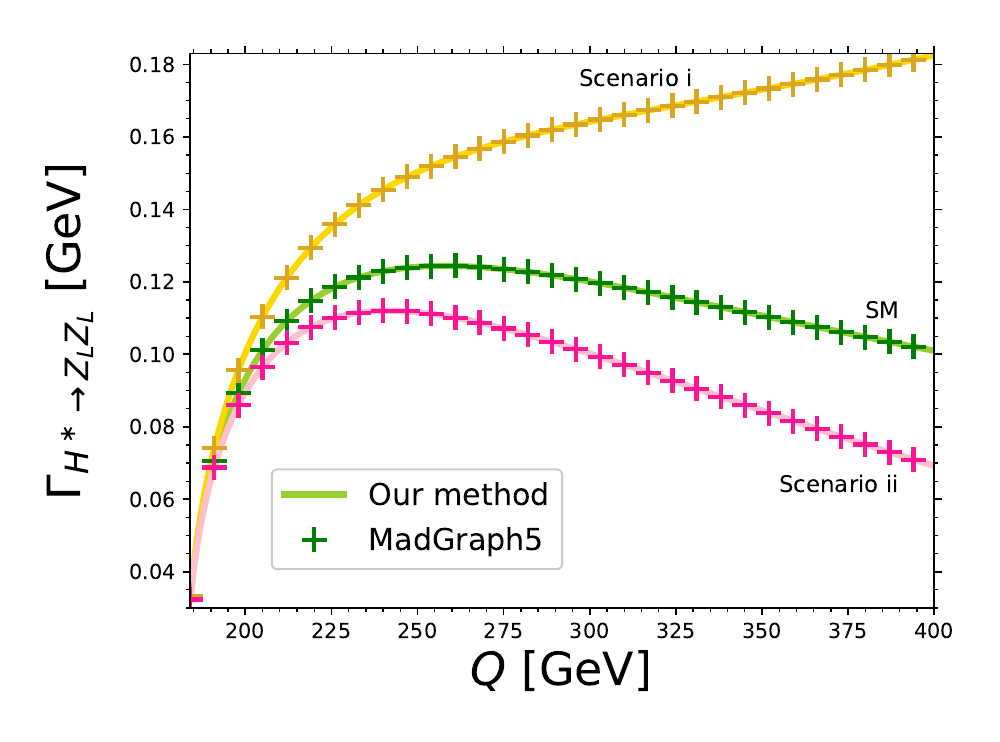}}\hspace{-.2cm}
{\includegraphics[width=9cm]{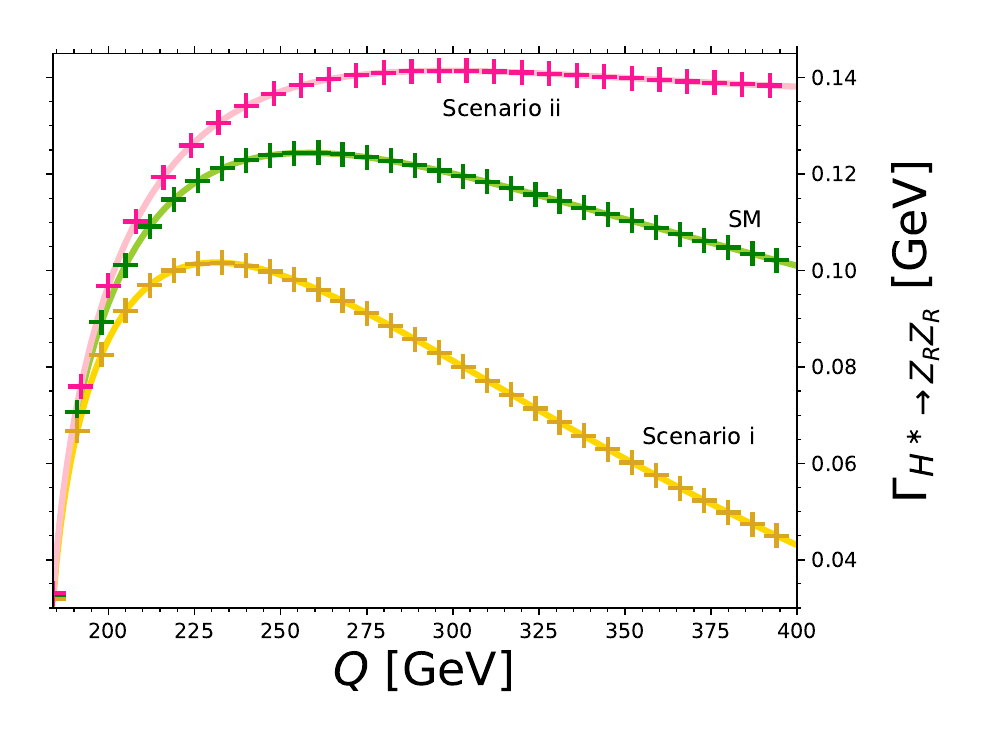}}
\caption{Behavior of the  polarized $H^\ast\rightarrow Z_LZ_L$ (left plot) and  $H^\ast\rightarrow Z_RZ_R$ (right plot) decay widths as functions of the four-lepton invariant mass for two sets of values of the anomalous couplings: scenario i)$\hat{b}_Z=0.001+i 0.003$, $\hat{c}_Z=0.001+i0.002$, and $\widetilde{b}_Z=0.01+i0.02$;  scenario ii)$\hat{b}_Z=0.0001+i 0.001$, $\hat{c}_Z=0.0001+i0.001$, and $\widetilde{b}_Z=0.001-i0.01$. We also show the SM tree-level contributions for comparison purposes. For the numerical evaluation we use our own evaluation method and \texttt{MadGraph5\_aMC@NLO}.  \label{poldecwidth}} 
\end{center}
\end{figure}

We now turn to cross-check the numerical results for the polarized  $H^\ast\rightarrow ZZ$ decay width, which is sensitive to the anomalous $HZZ$ couplings and yields  the polarized $H^\ast\rightarrow ZZ \rightarrow e^-e^+\mu^-\mu^+$ decay  via Eq. \eqref{TotWpol}.  For our numerical evaluation method, we use the analytical result of Eq. \eqref{HZZpolG} for the polarized partial decay widths $\Gamma^\lambda_{H^\ast\rightarrow Z_\lambda Z_\lambda}$. 
We  show in the left (right) plots of Fig. \ref{poldecwidth} the behavior of the $H^\ast\rightarrow Z_LZ_L$ ($H^\ast\rightarrow Z_R Z_R$) decay width as a function of the four-lepton invariant mass for two sets of values of the anomalous $HZZ$ couplings consistent with the most stringent bounds \cite{Hernandez-Juarez:2023dor}.   For comparison purposes, we also show the tree-level SM contributions.  

Again, we can conclude that the results obtained by our numerical evaluation method and \texttt{MadGraph5\_aMC@NLO} agree nicely.
We observe that the anomalous contributions yield a significant deviation from the tree-level SM contribution for large  $Q$, and also note that the $H^\ast\rightarrow Z_\lambda Z_\lambda$  decay width shows a distinctive behavior for each polarization in both scenarios of new physics. 
As discussed below, this stems from  the $CP$-violating effects and complex form factors, which have already been studied through a left-right asymmetry in the decay $H^\ast\rightarrow Z_\lambda Z_\lambda$  \cite{Hernandez-Juarez:2023dor}. A more detailed analysis of the polarized  $H^*\to \bar{\ell}_1 \ell_1 \bar{\ell}_2 \ell_2$  decay widths, including the SM contribution up to one-loop level, together with the analysis of other interesting observables sensitive to new physics will be presented below.

In summary, there is an excellent agreement between the results obtained by our  evaluation method and \texttt{MadGraph5\_aMC@NLO}.  We also corroborate the consistency of  Eqs.  \eqref{TotW} and \eqref{TotWpol}, which allows one to study the four-lepton decay widths using the $H^*\to ZZ$ decay width only. Note that in our evaluation method, we consider a constant $\Gamma_Z$ decay width in the narrow width approximation, whereas  in \texttt{MadGraph5\_aMC@NLO} the complex mass scheme is used. According to Ref.  \cite{Denner:2005fg},  both methods yield the same results for four-leptons in the final state. In the following, we will use our evaluation method  as it is more suited for our analysis.

\subsection{Polarized  decay widths}\label{InvMassDis}

 The physics of polarizations of weak bosons at the LHC has been discussed in Refs. \cite{Ballestrero:2019qoy,Maina:2020rgd,Maina:2021xpe}, whereas those related to the $Z$ boson are of particular interest as they are being measured by the CMS, ATLAS and LHCb collaborations \cite{CMS:2015cyj,ATLAS:2016rnf,LHCb:2022tbc,ATLAS:2022oge,ATLAS:2023zrv}. To study the polarized $Z$ boson effects,  we now consider the most general case with complex $HZZ$ anomalous couplings and examine three realistic scenarios for their numerical values. For the $\hat{b}_Z$ anomalous coupling, we will consider the SM contribution up to the one-loop level. That result has been reported in Ref. \cite{Hernandez-Juarez:2023dor} in terms of the Passarino-Veltman scalar functions and as a function of $Q$, the four-lepton invariant mass. The numerical evaluation of $\hat{b}_Z$ is obtained through the LoopTools package \cite{Hahn:1998yk}. While, for the $\hat{c}_Z$ and $\widetilde{b}_Z$ anomalous couplings, we use the stringent limits obtained through LHC data and theoretical results \cite{CMS:2022ley,Hernandez-Juarez:2023dor}. They are shown in Table \ref{scenarios}. Note that although we consider a non-vanishing $\hat{c}_Z$, it yields a negligible contribution to the $h_1^H$ form factor, and hence the same values are considered for the three scenarios. Finally, the one-loop SM contribution is also studied to compare with the new physics scenarios.   

\begin{table}[!htb]
\caption{Scenarios for the $HZZ$ anomalous couplings used in our analysis of the $H^\ast\rightarrow e^-e^+\mu^- \mu^+$ decay. For the $\hat{b}_Z$ coupling, we consider  SM contribution up to one-loop level \cite{Hernandez-Juarez:2023dor}, whereas for the remaining anomalous couplings, the stringent limits obtained through LHC data and theoretical results are used. \label{scenarios}}
\begin{tabular}{ccc}
\hline\hline
Scenario&$\widetilde{b}_Z$&$\hat{c}_Z$\\
\hline\hline
1& $0.01+i0.02$&$0.0001+0.0003i$\\
2& $0.001-i0.01$&$0.0001+0.0003i$\\
3& $0.0001+i0.001$&$0.0001+0.0003i$\\
\hline
\hline
\end{tabular}
\end{table}

\begin{figure}[!hbt]
\begin{center}
{\includegraphics[width=9cm]{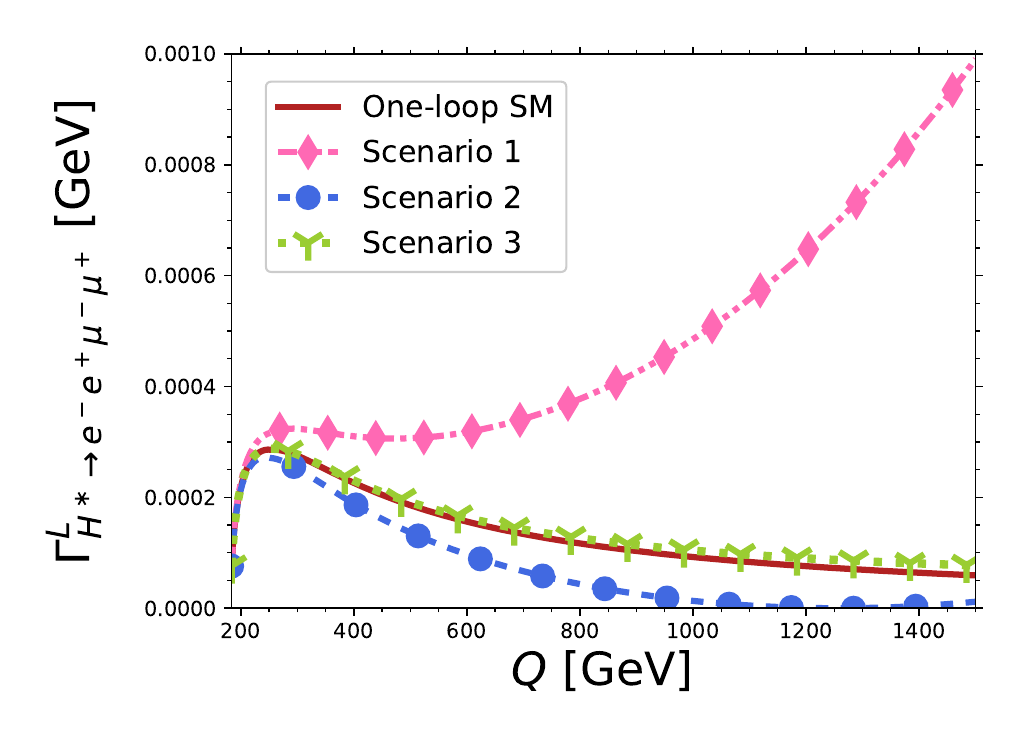}}\hspace{-.2cm}
{\includegraphics[width=9cm]{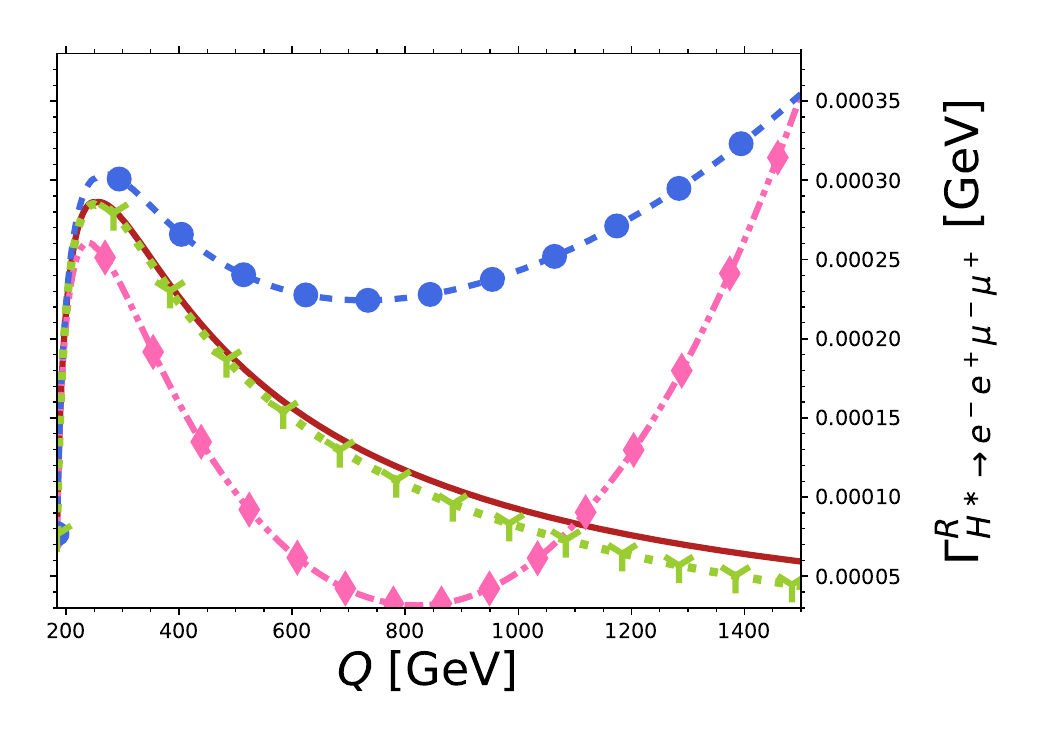}}
\caption{Behavior of the  transversally polarized $\Gamma^\lambda_{H^\ast\rightarrow e^-e^+\mu^- \mu^+}$ decay widths ($\lambda=L$, $R$)  for the SM contribution and the three scenarios of Table \ref{scenarios} for the anomalous $HZZ$ couplings.  \label{transpoldecwidth}} 
\end{center}
\end{figure}

We first analyze the effects of the $Z$ gauge bosons polarizations on the $H^\ast\rightarrow ZZ \rightarrow e^-e^+\mu^-\mu^+$ decay, for which we can use either Eq. \eqref{TotalWidthPol} or  Eq. \eqref{TotWpol} as they yield the same results. 
We show  in Fig. \ref{transpoldecwidth} the results for the left- and right-handed polarized $H^\ast\rightarrow e^-e^+\mu^- \mu^+$ decay widths as functions of $Q$. The longitudinally polarized $Z$ gauge bosons are not considered as for this case, the $H^\ast\rightarrow ZZ \rightarrow e^-e^+\mu^-\mu^+$ decay width is not sensitive to the $CP$-violating anomalous coupling $\tilde b_Z$.

We observe in Fig. \ref{transpoldecwidth} that for small values of the real and absorptive parts of the $CP$-violating anomalous coupling $\tilde b_Z$ (scenario 3), the polarized $\Gamma^\lambda_{H^\ast\rightarrow e^-e^+\mu^- \mu^+}$  decay widths shows a little variation from the pure SM contribution, though such a  deviation increases slightly as $Q$ grows, being more appreciable in the case of  right-handed polarization. On the other hand, in scenarios 1 and 2, in which $\tilde b_Z$ is assumed to be relatively large, a considerable deviation from the SM contribution develops around $Q=250$ GeV and  becomes larger as $Q$ increases.  Therefore,  the effects of the  complex anomalous $HZZ$ couplings in the four-lepton final state are more notable at high energies. Furthermore, in scenario 1,  $\Gamma_{H^\ast\rightarrow Z_LZ_L}$ is an increasing function of $Q$ and is larger than the tree-level SM contribution, whereas  $\Gamma_{H^\ast\rightarrow Z_RZ_R}$ decreases around $Q= 225$ GeV  and is smaller  than the tree-level SM contribution. For scenario 2 the contrary is true. Thus, an apposite behavior between the left and right polarized widths is observed.

Finally, in Fig. \ref{transpoldecwidth}, we note that the partial width increases significantly at high energies for some new physics scenarios, as  the non-SM contributions in Eq. \eqref{MLL}-\eqref{MRR}  are proportional to powers of $Q$.  A similar behavior has been observed in trilinear neutral gauge bosons couplings \cite{Baur:1992cd,Gounaris:1999kf}, where the unitarity can be preserved if we generalize the $\widetilde{b}_Z$ and $\hat{c}_Z$ anomalous couplings as 
\begin{equation}\label{uniExp}
\hat{c}_Z(Q^2)=\frac{\hat{c}_{Z0}}{\left(1+Q^2/\Lambda^2\right)^n}\text{,}\quad\quad \widetilde{b}_Z(Q^2)=\frac{\widetilde{b}_{Z0}}{\left(1+Q^2/\Lambda^2\right)^n},
\end{equation}
where $\hat{c}_{Z0}$ and $\widetilde{b}_{Z0}$ are constants, which can be  determined by considering unitarity conditions of the $gg\rightarrow H^\ast\rightarrow ZZ$ process \cite{Baur:1987mt}, whereas the parameter $\Lambda$ is an energy scale introduced to avoid  unphysical results as $Q$ increases. It is noted  in Eq. \eqref{uniExp} that the $\hat{c}_Z(Q^2)$ and $\widetilde{b}_Z(Q^2)$ anomalous couplings are approximately constants for $Q\ll \Lambda$. Therefore, our results in Fig. \ref{transpoldecwidth} are valid for an energy scale $\Lambda\gg$ 1 TeV, since we have considered the values  in Table. \ref{scenarios} and  there is no significant change in  the one-loop SM contribution \cite{Hernandez-Juarez:2023dor}. In general, the parameter $\Lambda$ is fixed to the order of TeVs or $\infty$ \cite{Baur:1992cd,ParticleDataGroup:2022pth}. For the latter case, the results obtained in this work remain unchanged. Nevertheless, for $\Lambda\sim$ 1 TeV the size of our predicted effects will reduce. The values of $\Lambda$ and $n$ still have to be determined for the $HZZ$ anomalous couplings, but this is beyond the scope of this work

\subsection{Left-right asymmetry $\mathcal{A}_{LR}$}
The behavior of the $H^\ast\rightarrow e^-e^+\mu^- \mu^+$ polarized amplitudes  give rise to a left-right asymmetry $\mathcal{A}_{LR}$
\begin{align}
\label{ALRe0}
\mathcal{A}_{LR}&=\frac{\Gamma^L_{H^\ast\rightarrow e^-e^+\mu^- \mu^+}-\Gamma^R_{H^\ast\rightarrow e^-e^+\mu^- \mu^+}}{\Gamma^L_{H^\ast\rightarrow e^-e^+\mu^- \mu^+}+\Gamma^R_{H^\ast\rightarrow e^-e^+\mu^- \mu^+}},
 \end{align}
which is non-vanishing as long as complex  anomalous couplings are present and can be written as 
\begin{align}
\label{ALRe}
\mathcal{A}_{LR}&= \frac{4  m_Z^2 \sqrt{Q^2\left(Q^2-4 m_Z^2\right)} {\rm Im}\left(h_3^H{h_1^H}^\dagger \right)}{Q^2
   \left(Q^2-4 m_Z^2\right)
   \big|h_3^H\big|^2+4 m_Z^4\big|h_1^H\big|^2},
   \end{align}
thereby being suitable for detecting $CP$-violating effects.

\begin{figure}[!hbt]
\begin{center}
\includegraphics[width=12cm]{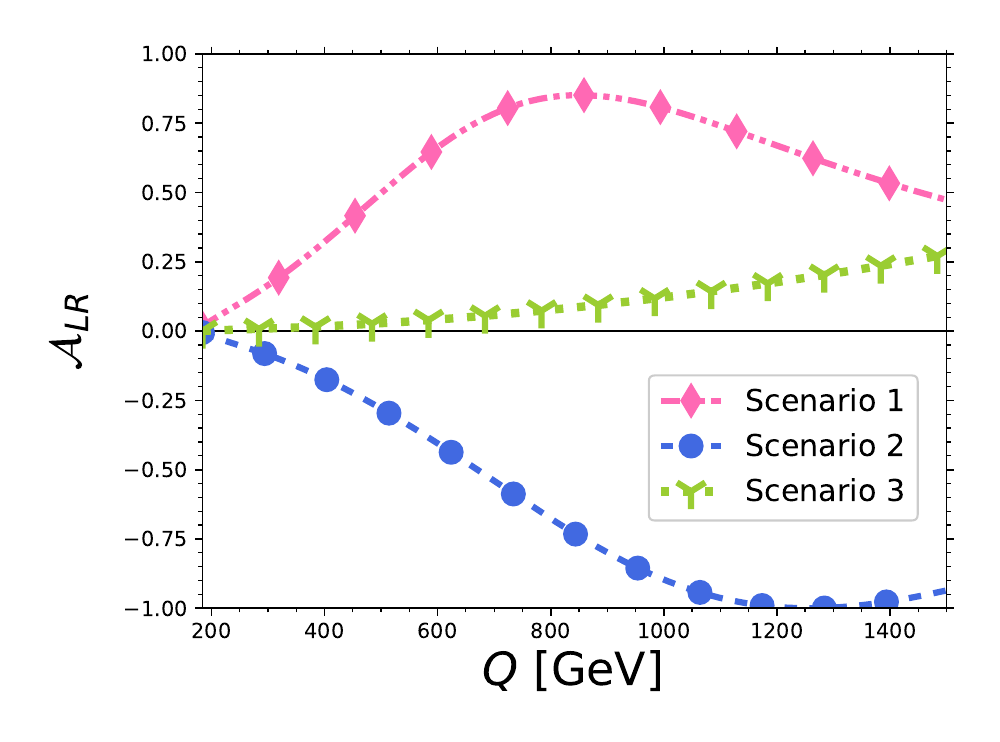}
\caption{Left-right asymmetry $\mathcal{A}_{LR}$ as a function of the four-lepton invariant mass in the three scenarios of Table \ref{scenarios} for the anomalous $HZZ$ couplings.   \label{FigLRA}} 
\end{center}
\end{figure}

In Fig. \ref{FigLRA}, we show the behavior of $\mathcal{A}_{LR}$ as a function of the four-lepton invariant mass in the three scenarios of Table \ref{scenarios} for the anomalous $HZZ$ couplings. The SM contribution is not shown since it vanishes at the one-loop level. We observe that for scenario 2, in which the real and absorptive parts of   $h_3^H$ are of opposite sign, the  $\mathcal{A}_{LR}$ magnitude can be larger than for the other cases. Also, it is noted that in the three scenarios, $\mathcal{A}_{LR}$ reaches its largest magnitude  at a relatively high $Q$, and hence the $CP$-violating effects would be more significant beyond the $2m_Z$ threshold. Nevertheless, the asymmetry tends to decrease  as $Q$ grows, which stems from the fact that it behaves as a $1/Q^2$ function at high energies.

The results of Figs. \ref{transpoldecwidth} and \ref{FigLRA}, together with Eq. \eqref{ALRe}, are in agreement with those reported in Ref. \cite{Hernandez-Juarez:2023dor} for the $H^\ast\rightarrow Z_\lambda Z_\lambda$ process, which is a consequence of  the four-leptons final state being a function of $Q$ and independent of the $Z\rightarrow \overline{\ell}\ell$ processes (see Eq. \ref{TotWpol}). On the other hand, the angular variables only appear in the production of lepton pairs and  could also serve to detect any new physics effects from to the $HZZ$ anomalous couplings.

\subsection{Angular distributions}\label{SecAngD}

\begin{figure}[!hbt]
\begin{center}
\includegraphics[width=7cm]{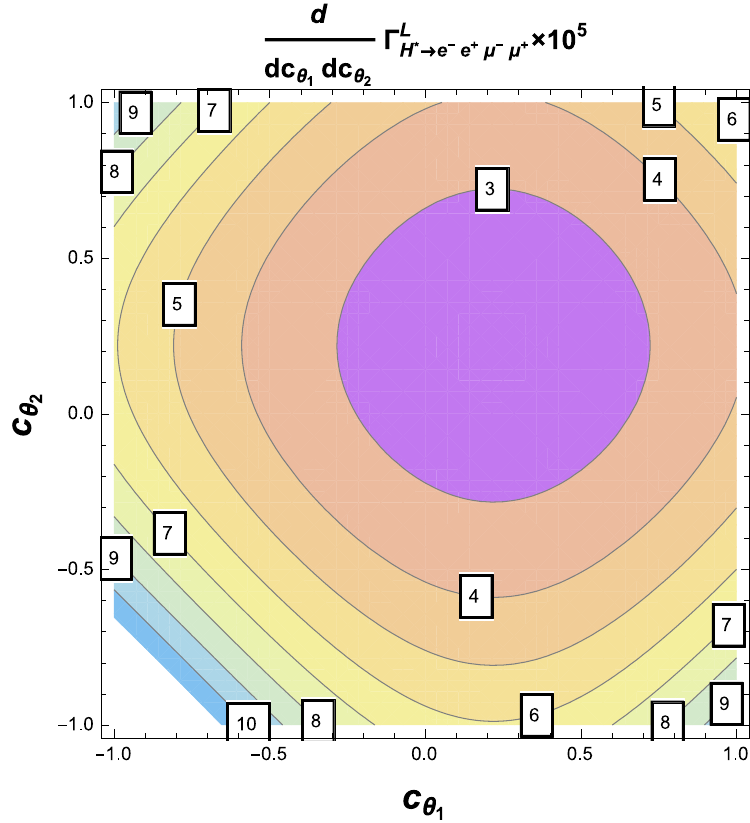}\hspace{-.01cm}
\includegraphics[width=7cm]{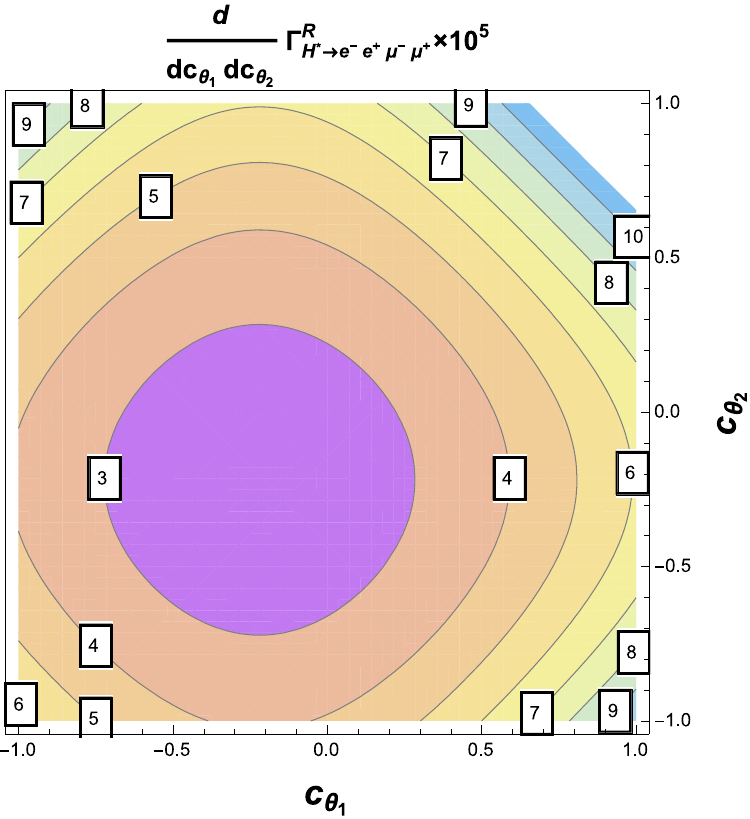}\hspace{-.01cm}
\caption{Contours of the transversally polarized $H^\ast\rightarrow e^-e^+\mu^- \mu^+$  differential decay widths in the $c_{\theta_1}$ vs $c_{\theta_2}$ plane for $Q=500$ GeV. We only include the SM one-loop contribution. \label{doubleD}} 
\end{center}
\end{figure}

The Higgs decay to four lepton is the cleanest channel to access to the polarizations of the weak bosons, as they can be studied through the angular observables of the leptons produced by the $Z$ bosons decays \cite{Ballestrero:2019qoy,Maina:2021xpe}. For that reason, we now turn to analyze the role of the angular variables in the polarized $H^\ast\rightarrow e^-e^+\mu^- \mu^+$  decay widths.  These angular distributions together with the $Z$ bosons polarizations are being measured at the LHC \cite{CMS:2015cyj,ATLAS:2016rnf,CMS:2018mbt,LHCb:2022tbc,ATLAS:2023lsr}. As we pointed out, the results obtained in this work remain unchanged for an energy scale $\Lambda\gg1$ TeV. Usually values for $\Lambda$ of this order are considered to set limits on similar anomalous couplings to those considered in the $HZZ$ interaction \cite{ParticleDataGroup:2022pth}. Since the angle $\phi$  only enters into the interference term of the full square amplitude, its effects  are unobservable through polarized $Z$ gauge bosons. After  the integration of Eq. \eqref{difEq2} over $\phi$,  the differential $H^\ast\rightarrow e^-e^+\mu^- \mu^+$ decay width will be  in terms of the angles $\theta_{1,2}$. Therefore, it is still possible to study the polarization effects in the four lepton-final state through the angular variables. From Eq. \eqref{MZLR1} we observe that the left- and right-handed square amplitudes  are symmetric in both $\theta_{1}$ and $\theta_{2}$  angles, which is evident  in Fig. \ref{doubleD}, where we show the contours of the SM contribution to the transversally polarized $H^\ast\rightarrow e^-e^+\mu^- \mu^+$ differential decay widths in the $c_{\theta_1}$ vs $c_{\theta_2}$ plane for $Q=500$ GeV.  
The left-handed polarized  case reaches its higher values around $\cos\theta_i=-1$, whereas for the right polarization this occurs at $\cos\theta_i=1$. In these regions the partial widths can be of order $10^{-4}$. Future experimental searches should focus on these specific angular regions to increase the sensitivity to potential deviations from the SM predictions. The smallest values are obtained around $\cos\theta_i\approx0$ in both scenarios. Furthermore, it is found that the transversally polarized differential decay widths for  $H^\ast\rightarrow e^-e^+\mu^- \mu^+$ exhibit similar behavior for other values of $Q$ and when non-zero anomalous couplings are considered. 


\begin{figure}[!hbt]
\begin{center}
\includegraphics[width=9.2cm]{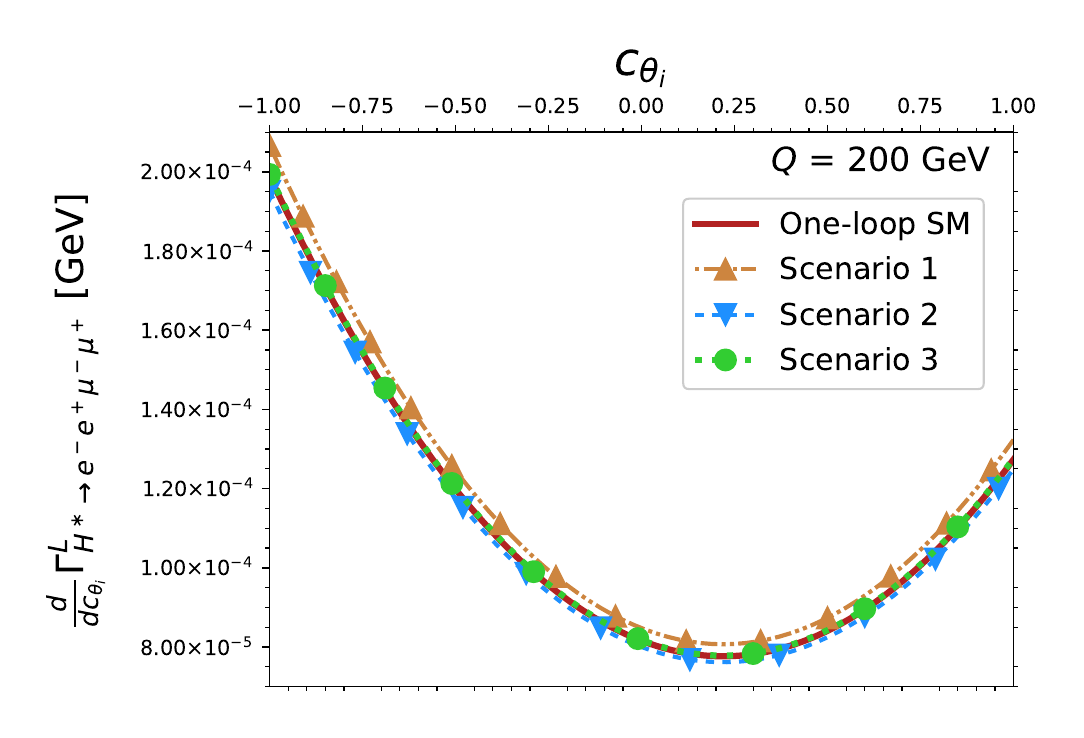}\hspace{-.73cm}
\includegraphics[width=9.3cm]{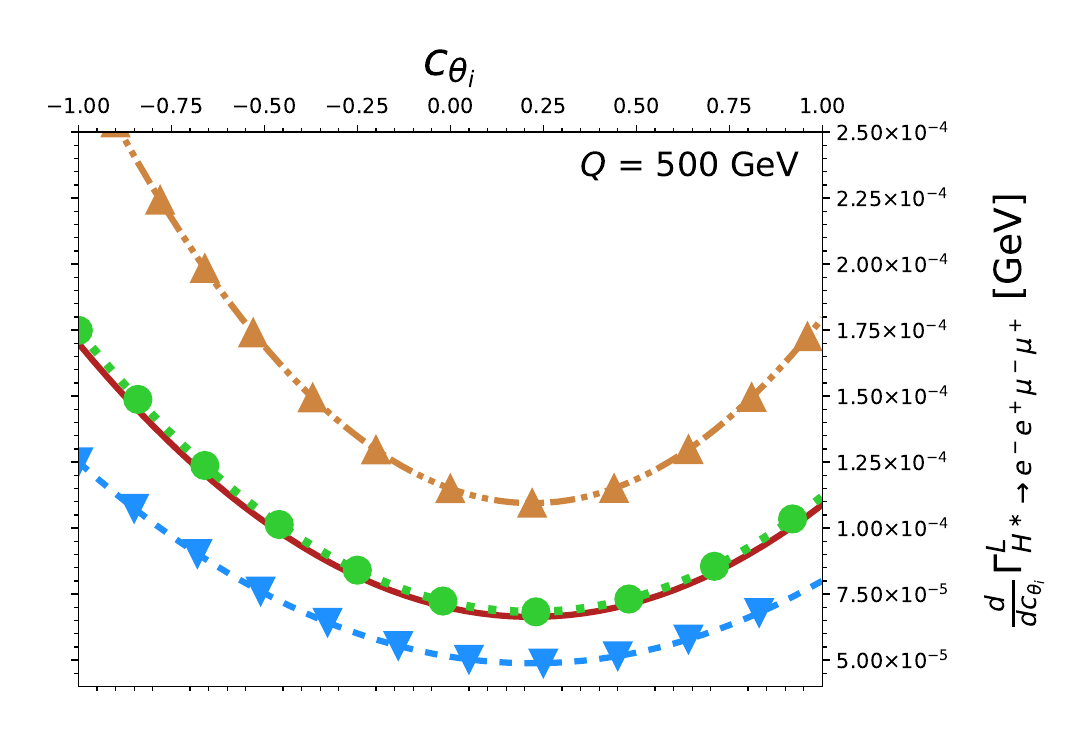}\\\vspace{-.78cm}
\includegraphics[width=9.25cm]{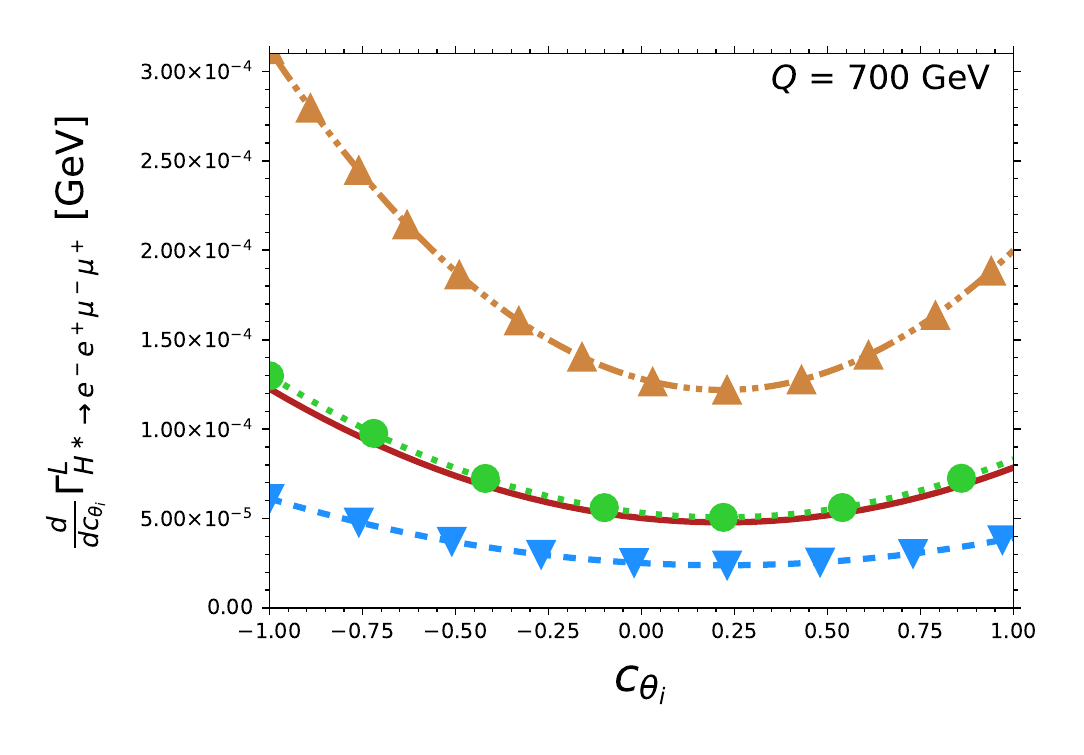}\hspace{-.73cm}
\includegraphics[width=9.25cm]{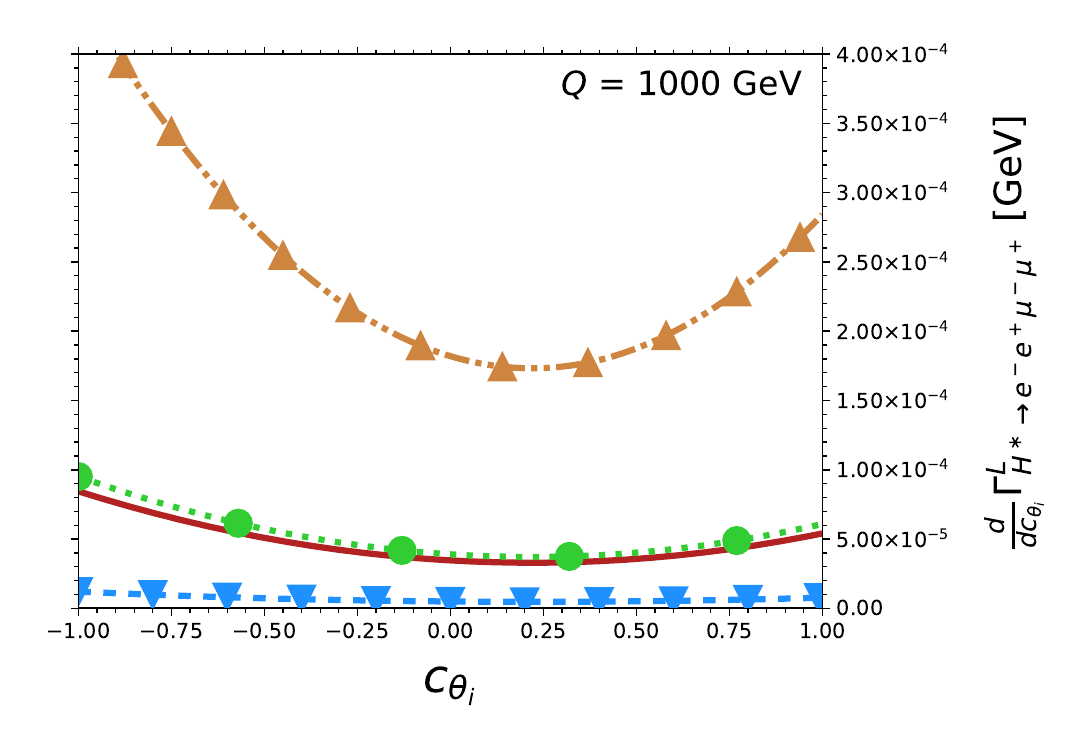}
\caption{Differential $H^\ast\rightarrow e^-e^+\mu^- \mu^+$  decay width for left-handed polarization as a function of $\cos\theta_i$ ($i=1$, 2) for a few values of the four-lepton invariant mass $Q$ in the three scenarios of Table \ref{scenarios} for the anomalous $HZZ$ couplings. The SM contribution up to the one-loop level is included. We already integrated over one of the two angles.  \label{4plotsL}} 
\end{center}
\end{figure}

\begin{figure}[!hbt]
\begin{center}
\includegraphics[width=9.2cm]{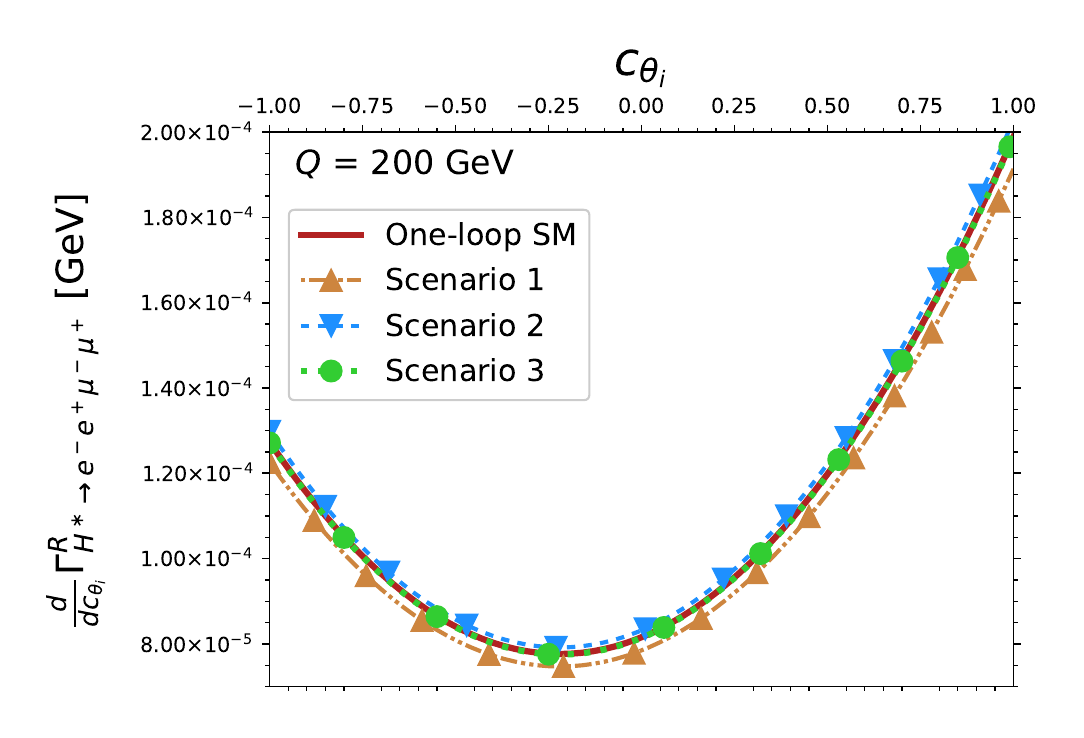}\hspace{-.73cm}
\includegraphics[width=9.3cm]{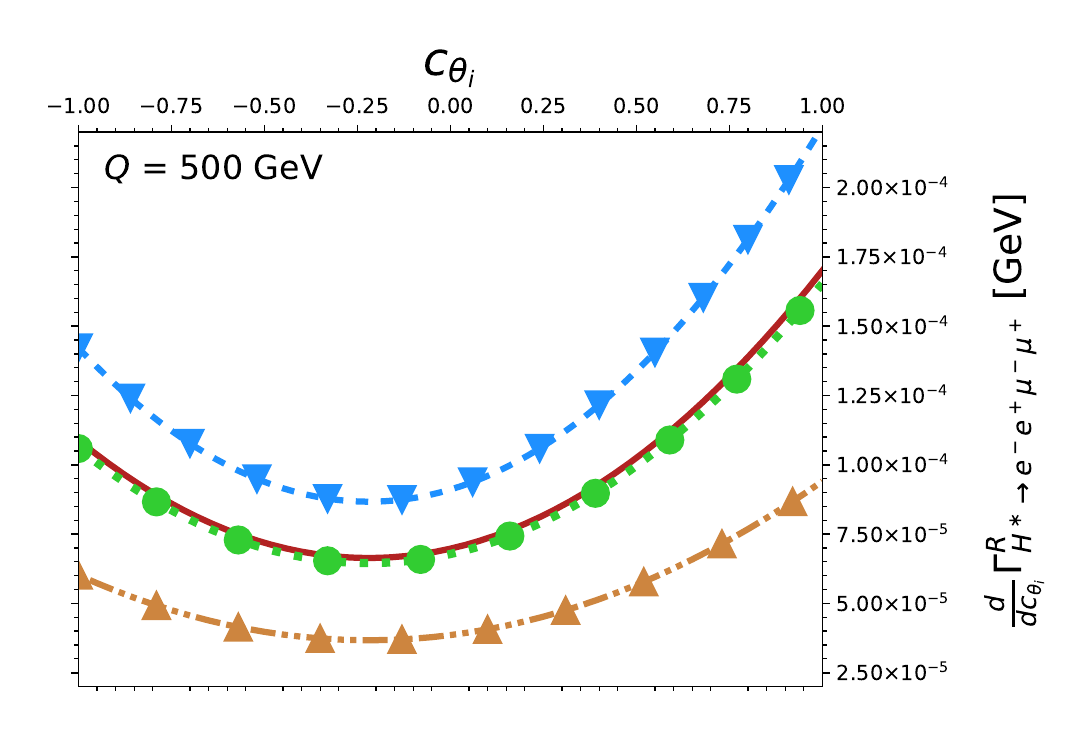}\\\vspace{-.78cm}
\includegraphics[width=9.25cm]{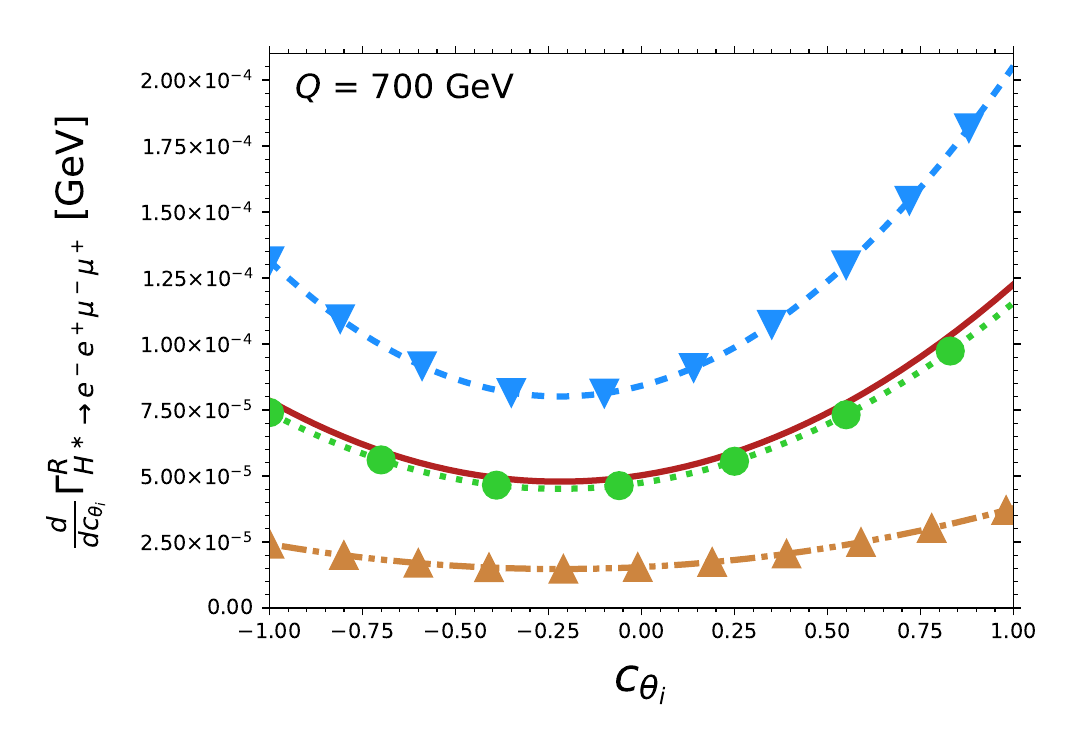}\hspace{-.73cm}
\includegraphics[width=9.25cm]{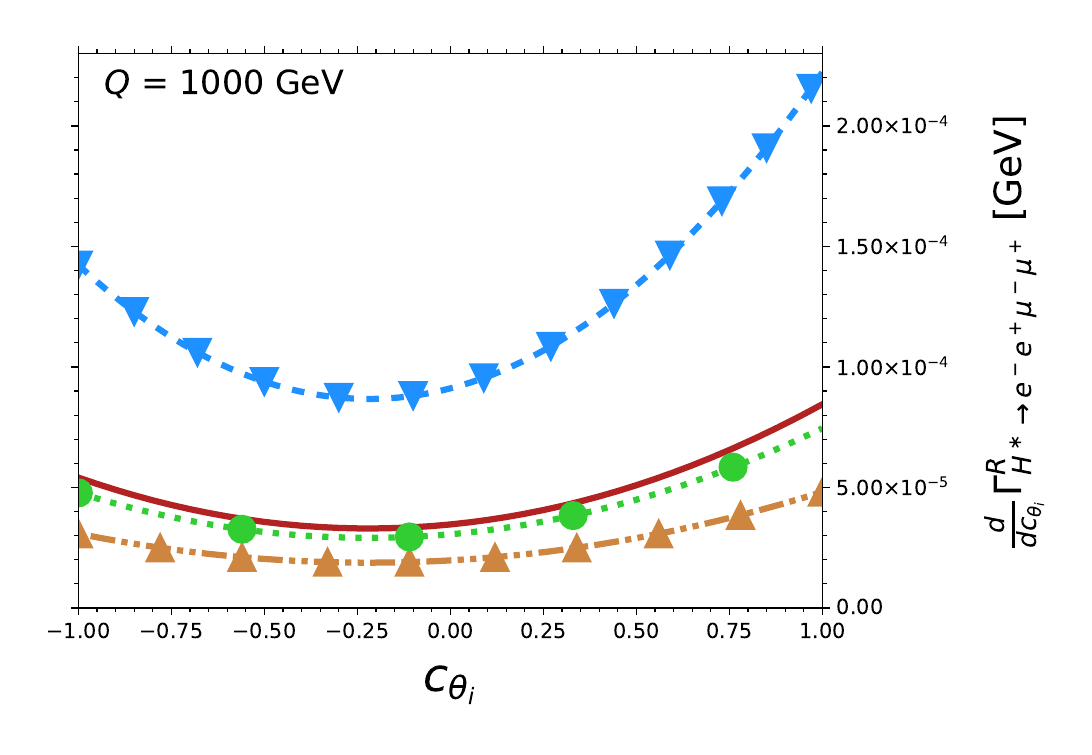}
\caption{The same as in Fig.  \ref{4plotsL}, but for right-handed polarizations.\label{4plotsR}} 
\end{center}
\end{figure}

We now consider the three scenarios of Table \ref{scenarios} for the anomalous $HZZ$ couplings and study their effects  on the polarized $H^\ast\rightarrow e^-e^+\mu^- \mu^+$  angular distributions after integrating over $\phi$ and one of the angles $\theta_{1,2}$. The results are presented in Figs. \ref{4plotsL}  and \ref{4plotsR} as functions of the cosine of the remaining angle $c_{\theta_i}$.
It is observed that  the angular distributions show a significant deviation from the SM contribution in scenarios 1 and 2, which becomes more pronounced at large $Q$. However, in scenario 3, where small values for the real and absorptive parts of the anomalous $h_3^H$ coupling are considered, the deviation  is insignificant for small $Q$, though it becomes noticeable as $Q$ becomes large, particularly in the case of right-handed polarization. We also note that, for left-handed  polarization, the differential decay width  is always above (below) the SM contribution in scenario 1 (scenario 2), whereas the opposite is true for right-handed polarization. 
Also,  the left-handed (right-handed) polarized angular distributions reach their larger magnitude as $c_{\theta_i}\to -1$ $(c_{\theta_i}\to 1)$. 
This distinctive behavior of the angular distributions for each transverse polarization of the $Z$ gauge bosons stems from the fact that the terms that give rise to the left-right asymmetry $\mathcal{A}_{LR}$ of Eq. \eqref{ALRe}  and  the terms  proportional to $c_{\theta_i}$ in Eq. \eqref{MZLR1}, which have opposite signs, remain unchanged after $\phi$ integration. This hints at  new asymmetries associated with the angular variables, which we will analyze below. We also would like to point out that the results shown in Figs. \ref{4plotsL} and \ref{4plotsR} for the SM contributions agree with those reported in Ref. \cite{Maina:2020rgd}. Some technics to identify $Z$ bosons polarizations through angular variables at the LHC have been addressed in Refs. \cite{Rao:2020hel,Maina:2021xpe,Aguilar-Saavedra:2024whi}, whereas using polarized beams in the future ILC or different $e^+e^-$ colliders have been also studied in Refs. \cite{Hagiwara:1993sw,Rao:2023jpr,Li:2024mxw}. These approaches may be extended to observe the polarized angular distributions at high-energy colliders.

\subsubsection{Polarized asymmetries}

As some asymmetries from the $HZZ$ coupling can be large enough to give evidence of new physics at the LHC \cite{Godbole:2007cn} and  motivated by the behavior of the $H^\ast\rightarrow e^-e^+\mu^- \mu^+$  angular distributions for transverse polarizations, we define the following angular left-right asymmetry
\begin{align}
\label{LRangEq}
\mathcal{A}_{LR\theta}&=\dfrac{\dfrac{d}{dc_{\theta_i}}\Gamma^L_{H^\ast\rightarrow e^-e^+\mu^- \mu^+}-\dfrac{d}{dc_{\theta_i}}\Gamma^R_{H^\ast\rightarrow e^-e^+\mu^- \mu^+}}
{\dfrac{d}{dc_{\theta_i}}\Gamma^L_{H^\ast\rightarrow e^-e^+\mu^- \mu^+}+\dfrac{d}{dc_{\theta_i}}\Gamma^R_{H^\ast\rightarrow e^-e^+\mu^- \mu^+}},
   \end{align}
where $i=1$ (2) and the angle $\theta_2$ ($\theta_1$) was integrated over.
Since $\Gamma_{H^\ast\rightarrow e^-e^+\mu^- \mu^+}$ is symmetric under $\theta_i$ ($i=1$, 2), the same expression holds  for both angles. The analytical form, obtained through  Eq. \eqref{difEq2}, can be written as
\begin{align}\label{LRangEq2}
\mathcal{A}_{LR\theta}&=- \frac{4}{m_Z^2}\frac{f(Q^2, c_{\theta_i} )}{h(Q^2, c_{\theta_i} )} ,
   \end{align}
with the $f(Q^2, c_{\theta_i} )$ and $h(Q^2, c_{\theta_i})$ functions given as  
   \begin{align}\label{f}
f(Q^2, c_{\theta_i} )&=g_A g_V  c_{\theta_i} m_Z^2 \left(Q^2
   \left(Q^2-4 m_Z^2\right)\big|h_3^H\big|^2+4
   m_Z^4\big|h_1^H\big|^2\right)\nonumber\\&-
 \left(g_A^2 +g_V^2\right)  \left(1+c^2_{\theta_i}\right)\left(   m_Z^4 
   \sqrt{Q^2\left(Q^2-4 m_Z^2\right)}{\rm Im}\left(h_3^H{h_
   1^H}^\dagger\right)\right),
\end{align}
   \begin{align}
h(Q^2, c_{\theta_i} )&=-16  g_A g_V
    c_{\theta_i} m_Z^2
   \sqrt{Q^2\left(Q^2-4 m_Z^2\right)}
  {\rm Im}\left(h_3^H{h_
   1^H}^\dagger\right)\nonumber\\&+\left(g_A^2+g_V^
   2\Big)   \left(1+c^2_{\theta_i}\right)\Big(Q^2
   \left(Q^2-4 m_Z^2\right)
   \big|h_3^H\big|^2+4
   m_Z^4 \big|h_1^H\big|^2\right),
\end{align}
where  the terms proportional to $g_{V,A}^2$ in $f\left(Q^2, c_{\theta_i} \right)$ are identical to those in the $\mathcal{A}_{LR}$ asymmetry Eq. \eqref{ALRe}, whereas the remaining terms arise from those proportional to $c_{\theta_i}$ in the $H^\ast\rightarrow e^-e^+\mu^- \mu^+$ amplitude, which have been not integrated out yet. Furthermore,  it is noted that if  the  numerator and denominator  Eq. \eqref{LRangEq2} are integrated over $c_{\theta_i}$ we obtain the non-angular $\mathcal{A}_{LR}$ asymmetry. Although the $\mathcal{A}_{LR}$ and $\mathcal{A}_{LR\theta}$  asymmetries are related, they have a distinct origin: the former is induced via the $HZZ$ anomalous couplings, and  the latter is a result of  polarized  $Z$ gauge bosons decaying into leptons, where the angular variables appear. It is also worth noting that even for a vanishing $CP$-violating form factor $h_3^H$, the $\mathcal{A}_{LR\theta}$ asymmetry is  non-zero, thereby being non-vanishing in the SM  at the one-loop level:
  \begin{align}
\mathcal{A}^{\rm SM}_{LR\theta}=-\frac{4 g_A g_V
    c_{\theta _i}}{
   \left( g_A^2+g_V^2\right)  \left(1+c
   ^2_{\theta
   _i}\right)}.\end{align}

\begin{figure}[!hbt]
\begin{center}
\includegraphics[width=9.2cm]{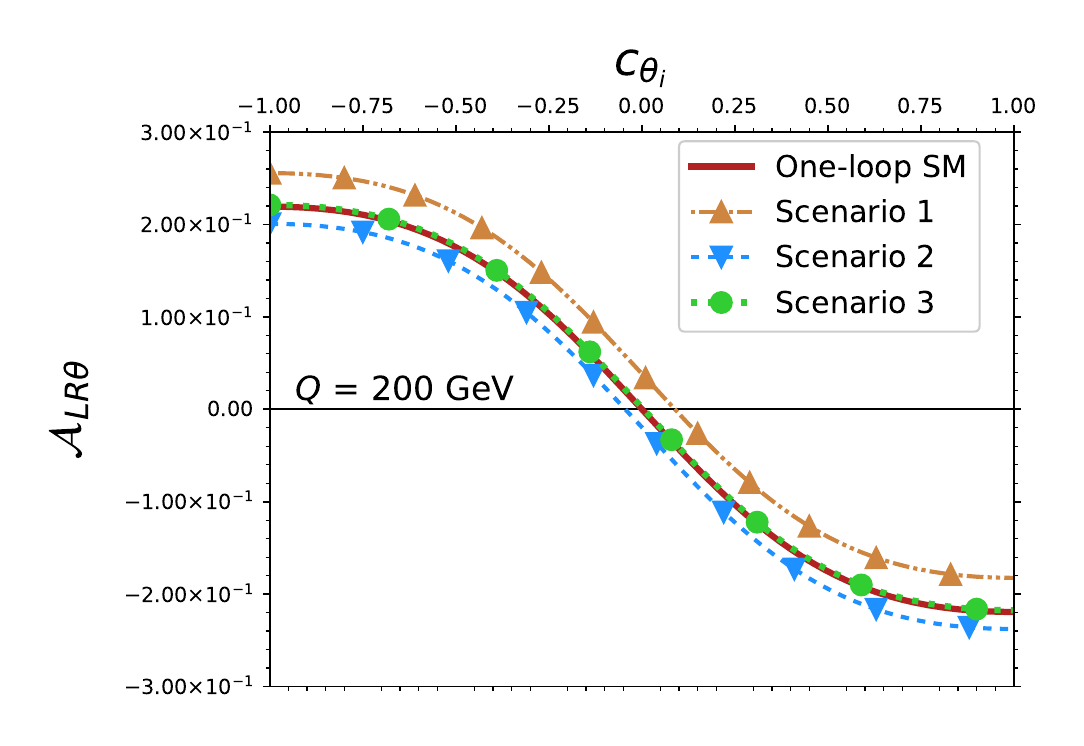}\hspace{-.73cm}
\includegraphics[width=9.3cm]{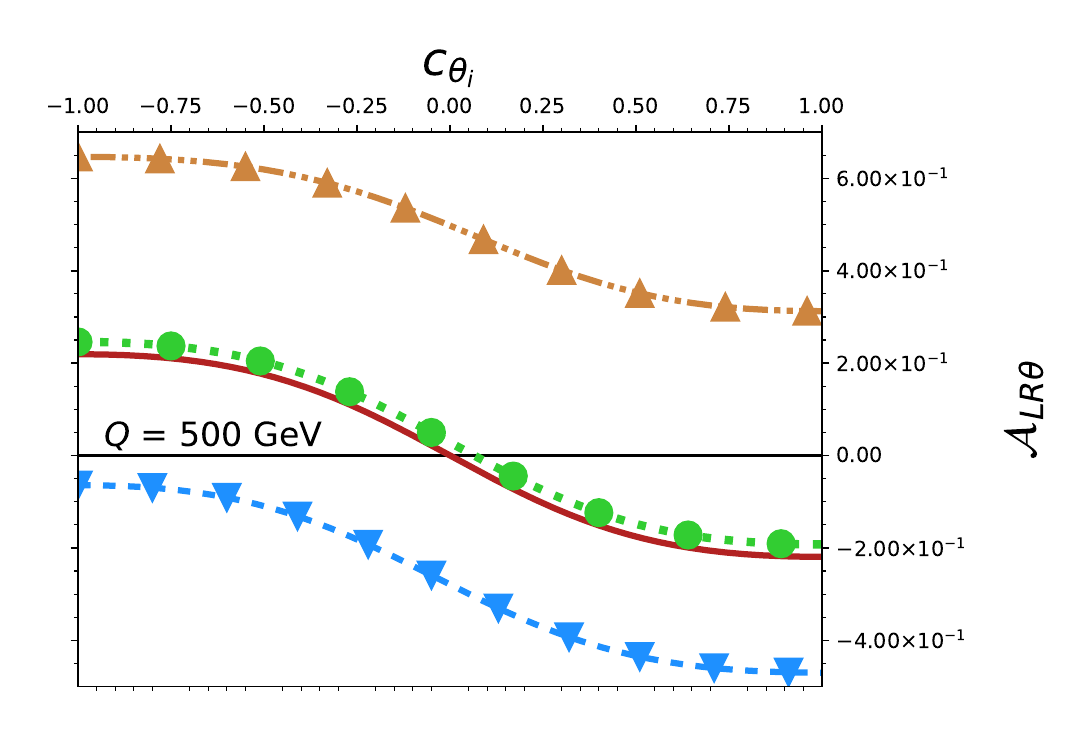}\\\vspace{-.78cm}
\includegraphics[width=9.25cm]{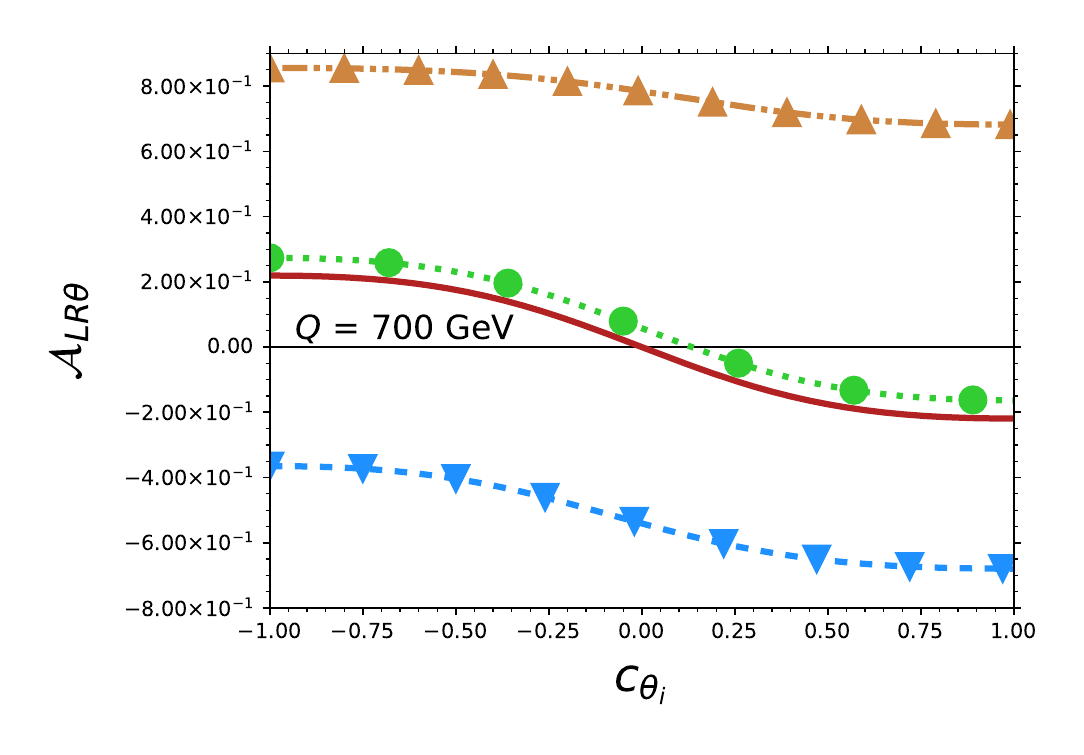}\hspace{-.73cm}
\includegraphics[width=9.25cm]{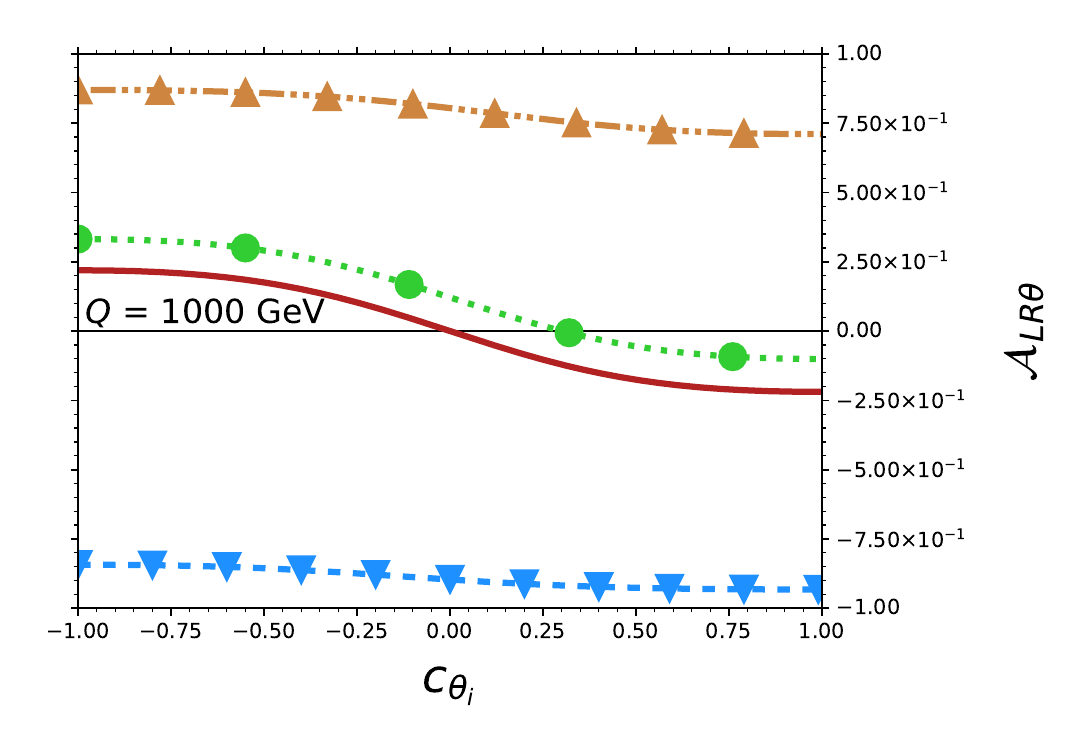}
\caption{Behavior of the left-right asymmetry $\mathcal{A}_{LR\theta}$ as a function $\cos\theta_i$ for a few values of  the four-lepton invariant mass $Q$ in the three scenarios of Table \ref{scenarios} for the anomalous $HZZ$ couplings.   We also include the SM contribution up to the one-loop level. \label{LRTasy}} 
\end{center}
\end{figure}

Again we consider the three scenarios of Table \ref{scenarios} for the anomalous $HZZ$ couplings and evaluate the effects on the  $\mathcal{A}_{LR\theta}$ asymmetry. The results are shown in Fig. \ref{LRTasy}, where  the SM contribution is also included. We observe that at low $Q$, $\mathcal{A}_{LR\theta}$ shows a slight deviation from the SM in scenarios 1 and 2, which becomes more significant as $Q$ increases, whereas such a variation is negligible in scenario 3. It is interesting to note that the deviation of $\mathcal{A}_{LR\theta}$  from the SM value is of the order of $10^{-1}-10^{-2}$, whereas  the corresponding difference of the polarized $H^\ast\rightarrow e^-e^+\mu^- \mu^+$ decay widths concerning the SM contribution is of the  order of $10^{-4}-10^{-5}$. Thus, there may be more possibilities to observe the effects of $CP$ violation through the $\mathcal{A}_{LR\theta}$ asymmetry  than in the polarized differential widths. 

Given the behavior observed in the polarized differential $H^\ast\rightarrow e^-e^+\mu^- \mu^+$  decay width, we also expect the presence of a polarized forward-backward asymmetry, which in general can be defined as follows for the polarized $H^\ast\rightarrow \overline{f}_if_i \overline{f}_jf_j$ decay
\begin{equation}
\mathcal{A}_{FB}^\lambda=\dfrac{
\displaystyle\int_{0}^1dc_{\theta_i}\dfrac{d}{dc_{\theta_i}}\Gamma^\lambda_{H^\ast\rightarrow \overline{f}_if_i \overline{f}_jf_j}-\int_{-1}^0dc_{\theta_i}\dfrac{d}{dc_{\theta_i}}\Gamma^\lambda_{H^\ast\rightarrow \overline{f}_if_i \overline{f}_jf_j}}
{\displaystyle\int_{0}^1dc_{\theta_i}\dfrac{d}{dc_{\theta_i}}\Gamma^\lambda_{H^\ast\rightarrow \overline{f}_if_i \overline{f}_jf_j}+\int_{-1}^0dc_{\theta_i}\dfrac{d}{dc_{\theta_i}}\Gamma^\lambda_{H^\ast\rightarrow \overline{f}_if_i \overline{f}_jf_j}}.
\end{equation}
A straightforward calculation yields 
\begin{align}\label{FBang}
\mathcal{A}_{FB}^{L/R}=\mp\frac{3 g_A g_V}{2
   \left(g_A^2+g_V^2\right)}.
\end{align}
Such a simple expression stems from the fact that $\mathcal{A}_{FB}^{L/R}$ is due to the $Z\rightarrow \overline{\ell}\ell$ decay, thereby being  independent  the anomalous $HZZ$ couplings and the four-lepton invariant mass: it is constant even in new-physics scenarios but a difference of sign arises for left-handed and right-handed polarizations. 

We present in Table \ref{tabAs} the values of the $\mathcal{A}_{FB}^\lambda$ asymmetry for all the 
combinations of light fermion flavors in the  $H^\ast\rightarrow \overline{f}_if_i \overline{f}_jf_j$ final state. We consider massless fermions and distinct fermion-antifermion pairs. It is also noted that the result of Eq. \eqref{FBang} is similar to the left-right $Z$ gauge boson asymmetry in the SM \cite{SLD:1994cex,Narita:1998rn}. 
\begin{table}[H]
\begin{center}  \caption{ $\mathcal{A}_{FB}^{L/R}$ asymmetry for all combinations of distinct pairs of light fermion-antifermion flavors in the  $H^\ast\rightarrow \overline{f}_if_i \overline{f}_jf_j$ decay. The fermion masses have been neglected.}\label{tabAs}
\begin{tabular}{ccc}
\hline \hline 
$f_i$&$f_j$& $\mathcal{A}_{FB}^{L/R}$  \\ 
\hline \hline 
$e$, $\mu$, $\tau$&$e$, $\mu$, $\tau$ &$\mp$0.164 \\
$\nu_e$, $\nu_\mu$, $\nu_\tau$&$\nu_e$, $\nu_\mu$, $\nu_\tau$ &$\mp$0.75 \\
$d$, $s$, $b$&$d$, $s$, $b$  &$\mp$0.705 \\
$u$, $c$ &$u$, $c$&$\mp$0.524 \\
\hline\hline \end{tabular} 
\end{center}
\end{table}

\subsubsection{Unpolarized asymmetries}

Up to now, we have focused only on the effects of the $Z$ gauge boson polarizations in the $H^\ast\rightarrow ZZ \rightarrow \overline{\ell}_1\ell_1\overline{\ell}_2\ell_2$ decay. Nevertheless,  the change  of sign in Eq. \eqref{FBang} hints the presence of a forward-backward asymmetry  $\mathcal{A}_{FB}$ in the unpolarized case, which is defined as follows
\begin{equation}
\mathcal{A}_{FB}=\dfrac{
\displaystyle\int_{0}^1dc_{\theta_i}\dfrac{d}{dc_{\theta_i}}\Gamma_{H^\ast\rightarrow ZZ \rightarrow e^-e^+\mu^-\mu^+}-\int_{-1}^0dc_{\theta_i}\dfrac{d}{dc_{\theta_i}}\Gamma_{H^\ast\rightarrow ZZ \rightarrow e^-e^+\mu^-\mu^+}}
{\displaystyle\int_{0}^1dc_{\theta_i}\dfrac{d}{dc_{\theta_i}}\Gamma_{H^\ast\rightarrow ZZ \rightarrow e^-e^+\mu^-\mu^+}+\int_{-1}^0dc_{\theta_i}\dfrac{d}{dc_{\theta_i}}\Gamma_{H^\ast\rightarrow ZZ \rightarrow e^-e^+\mu^-\mu^+}}.
\end{equation}
After inserting Eq. \eqref{difEq2} we obtain 
\begin{align}\label{FBunpol}
\mathcal{A}_{FB}=-\frac{48  g_A g_V
   m_Z^6
  Q \sqrt{Q^2-4 m_Z^2}}{\left(g_A^2+g_V^2\right)}
   \frac{{\rm Im}\left(h_3^H{h_
   1^H}^\dagger\right)}{G\left(Q^2\right)},
   \end{align}
where the  $G\left(Q^2\right)$ function was given in \eqref{Gfunc}.
We note that $\mathcal{A}_{FB}$ vanishes in the SM at the one-loop level since $h_3^H=0$. Our result  reproduces the one reported in  Ref. \cite{Godbole:2007cn}, where the absorptive part of $h_1^H$ was dismissed. However, such a term  is induced at the one-loop level in the SM, and its contribution provided that a non-vanishing new-physics contribution to $h_3^H$ is present, could be larger than that arising from the corresponding real part \cite{Hernandez-Juarez:2023dor}. Thus, our result is more complete than those previously reported in the literature. 
   
In contrast with the polarized forward-backward asymmetry, which is due only to the $Z \rightarrow \overline{\ell}_i\ell_i$ decay, the unpolarized one depends on the properties of both the $H^\ast\rightarrow ZZ$ and the $Z \rightarrow \overline{\ell}_i\ell_i$ decays: the dependence on the complex anomalous couplings  is a reflect of the former and the presence of the vector $g_V$ and axial $g_A$ couplings arises from the latter.
   
\begin{figure}[!hbt]
\begin{center}
{\includegraphics[width=12cm]{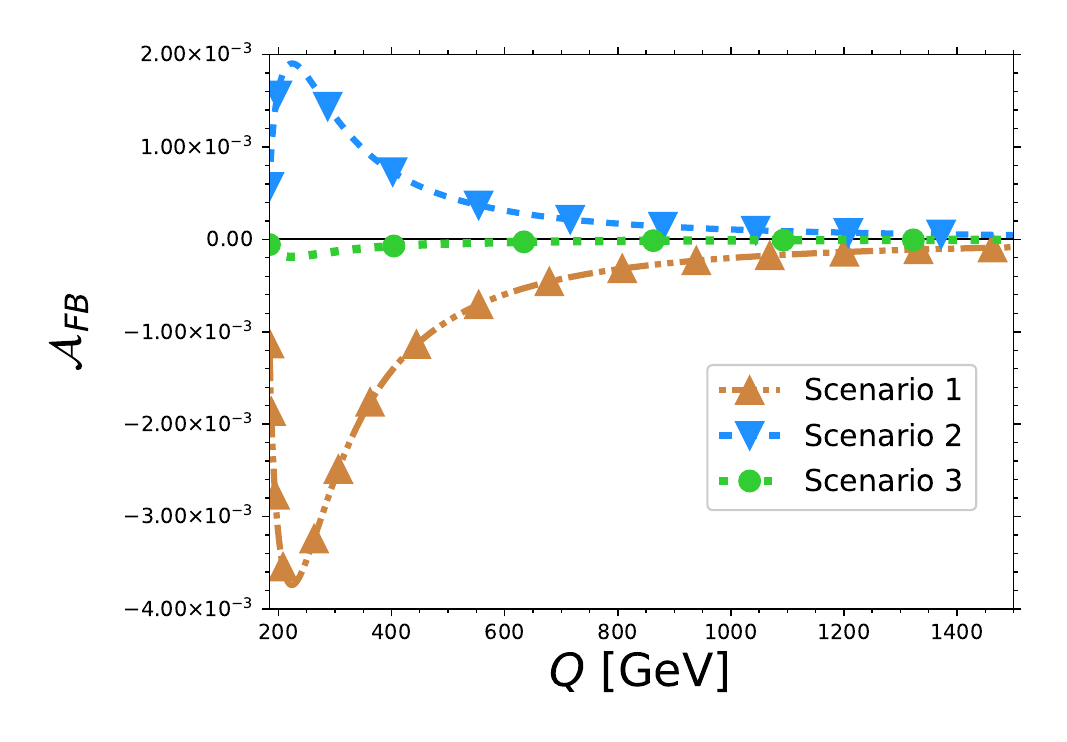}\label{plothhh7}}\hspace{-.2cm}
\caption{Behavior of the unpolarized forward-backward asymmetry $\mathcal{A}_{FB}$ as a function of the four-lepton invariant mass $Q$ in the three scenarios of Table \ref{scenarios} for the anomalous $HZZ$ couplings.  The one-loop SM contribution to the $h_1^H$ form factor has been added to the new-physics contribution.  \label{FBunpolFig}} 
\end{center}
\end{figure}

We show in Fig. \ref{FBunpolFig} the unpolarized forward-backward asymmetry $\mathcal{A}_{FB}$ as a function of the four-lepton invariant mass in the three scenarios of Table \ref{scenarios} for the anomalous $HZZ$ couplings. In addition to the anomalous contributions, we have included the one-loop SM contribution to the real and absorptive parts of the $h_1^H$ form factor. 
Similar to the above results for other observables,  $\mathcal{A}_{FB}$ can reach the largest values in scenarios 1 and 2,  being of the order of $10^{-3}$ at most, whereas it is one order of magnitude below in scenario 3. While $\mathcal{A}_{FB}$ can be large  at small energies, it tends to vanish at very large $Q$. An opposite behavior is observed for the $\mathcal{A}_{LR}$ and $\mathcal{A}_{LR\theta}$ asymmetries, which can reach values two orders of magnitude larger than that of $\mathcal{A}_{FB}$. To our knowledge,  the $\mathcal{A}_{FB}$ asymmetry has not been analyzed in the literature.

\subsubsection{$\phi$ distributions}

\begin{figure}[!hbt]
\begin{center}
\includegraphics[width=9.23cm]{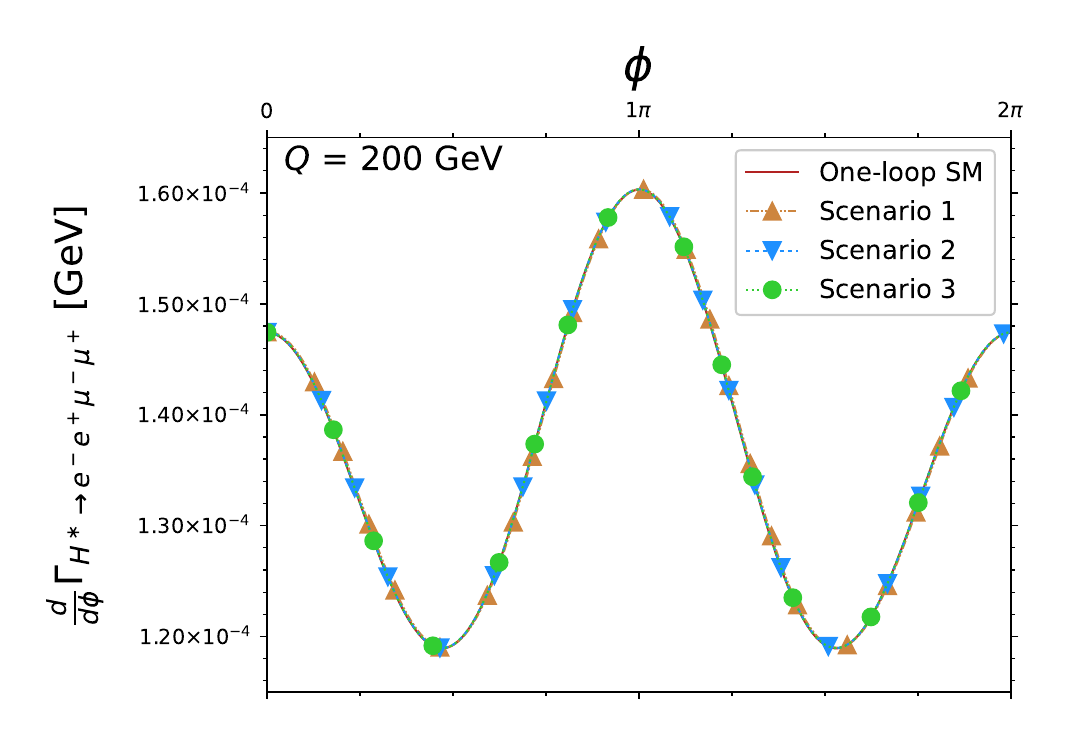}\hspace{-.73cm}
\includegraphics[width=9.19cm]{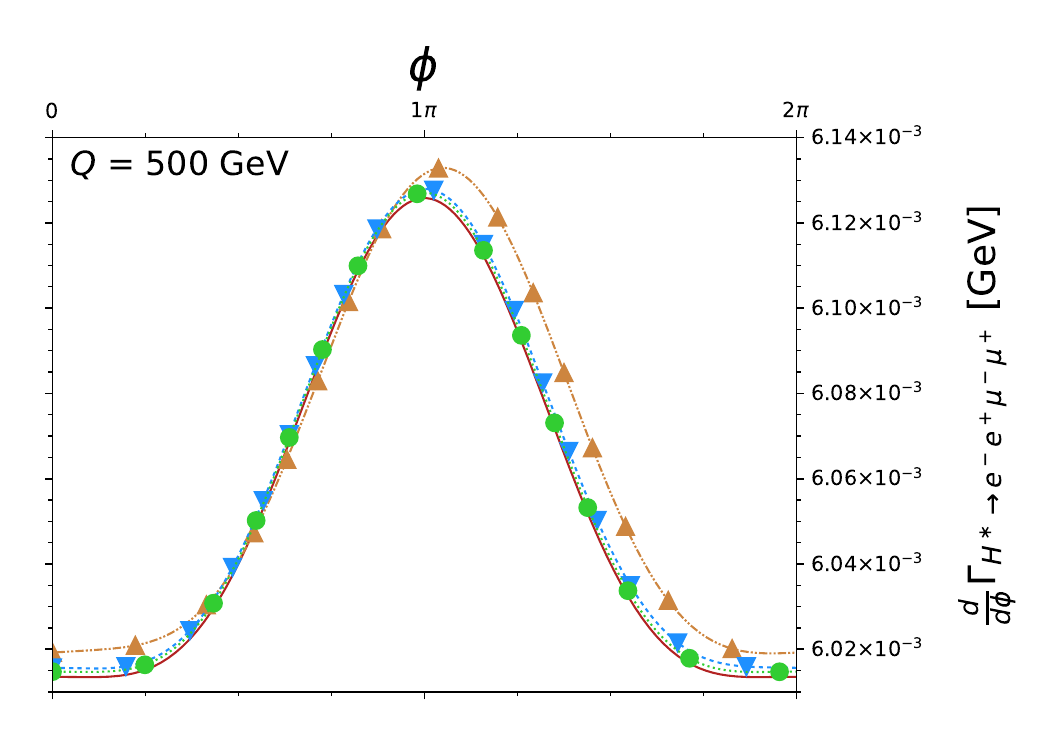}\\\vspace{-.78cm}
\includegraphics[width=9.25cm]{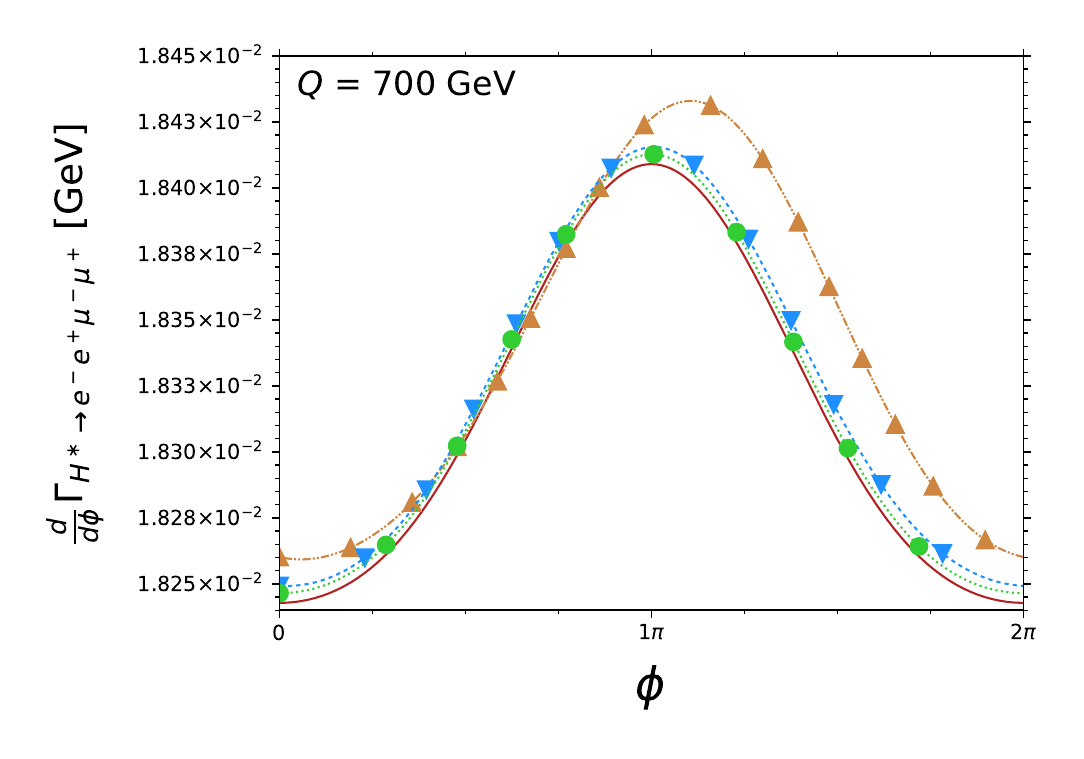}\hspace{-.73cm}
\includegraphics[width=8.9cm]{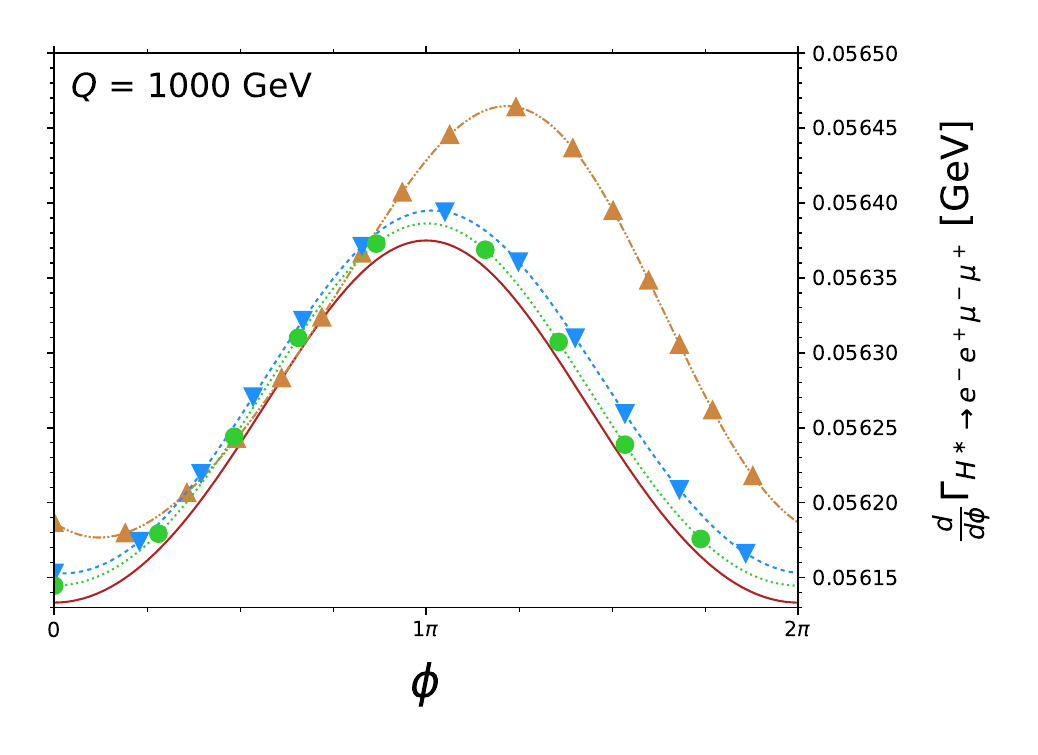}
\caption{Differential $H^\ast\rightarrow e^-e^+\mu^- \mu^+$  decay width  as a function of $\phi$  for various values of the four-lepton invariant mass $Q$ in the three scenarios of Table \ref{scenarios} for the anomalous $HZZ$ couplings. The SM contribution up to the one-loop level is included.\label{phiPlots}} 
\end{center}
\end{figure}

The effects of the anomalous couplings on $\phi$ distributions have been neglected up this point as they have been integrated out  in previous sections. These effects can be studied through the interference term $\mathcal{M}^2_{\text{int}}$ (Eq. \eqref{inter}) after integrating Eq. \eqref{difEq2}  over $c_{\theta_i}$ ($i$=1, 2). In Fig. \ref{phiPlots}, we show the numerical results of the $H^\ast\rightarrow e^-e^+\mu^- \mu^+$ partial width as a function of the azimuthal angle $\phi$ for the SM at one-loop level and the three scenarios in Table \ref{scenarios} for the $HZZ$ anomalous  couplings. It is observed that   larger values of $\Gamma_{H^\ast\rightarrow e^-e^+\mu^- \mu^+}$ are obtained at  high energies, where the new physics effects are also notable as significant deviations from the SM are found. Furthermore, a shift between the cases with non-SM anomalous couplings and the SM case is also observed. The consequences of the anomalous couplings are more pronounced for Scenario 1, which considers the largest values of the $CP$-violating form factor. 

The shift observed in the $\phi$ distributions leads to the definition of the $\mathcal{A}_\phi$ azimuthal asymmetry
\begin{equation}
\mathcal{A}_{\phi}=\dfrac{
\displaystyle\int_{\pi}^{2\pi}d\phi\dfrac{d}{d\phi}\Gamma_{H^\ast\rightarrow ZZ \rightarrow e^-e^+\mu^-\mu^+}-\int_{0}^\pi d\phi\dfrac{d}{d\phi}\Gamma_{H^\ast\rightarrow ZZ \rightarrow e^-e^+\mu^-\mu^+}}
{\displaystyle\int_{\pi}^{2\pi}d\phi\dfrac{d}{d\phi}\Gamma_{H^\ast\rightarrow ZZ \rightarrow e^-e^+\mu^-\mu^+}+\int_{0}^\pi d\phi\dfrac{d}{d\phi}\Gamma_{H^\ast\rightarrow ZZ \rightarrow e^-e^+\mu^-\mu^+}}.
\end{equation}

\begin{figure}[!hbt]
\begin{center}
{\includegraphics[width=12cm]{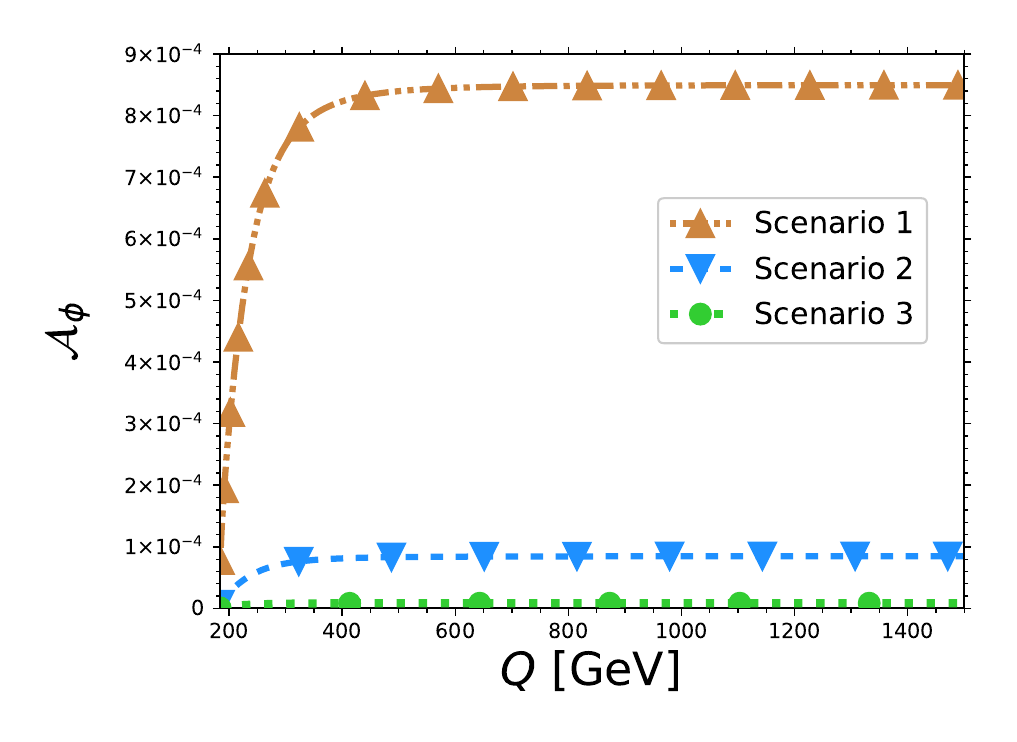}}\hspace{-.2cm}
\caption{Behavior of the azimuthal asymmetry $\mathcal{A}_{\phi}$ as a function of the four-lepton invariant mass $Q$ in the three scenarios of Table \ref{scenarios} for the anomalous $HZZ$ couplings.  The one-loop SM contribution to the $h_1^H$ form factor has been added to the new-physics contribution.  \label{plotAsyAz}} 
\end{center}
\end{figure}

After integrating Eq. \eqref{difEq2} we obtain 
\begin{align}\label{ASazi}
\mathcal{A}_{\phi}=-\frac{9\  \pi\ g^2_A g^2_V
   m_Z^2
  Q \sqrt{Q^2-4 m_Z^2}}{2\left(g_A^2+g_V^2\right)^2}
   \frac{K\left(Q^2\right)}{G\left(Q^2\right)},
   \end{align}
with 
\begin{align}\label{Kfunction}
K\left(Q^2\right)=&-2m_Z^2Q^2\Big[{\rm Re}\left(h_1^H{h_
   3^H}^\dagger\right)-2{\rm Re}\left(h_2^H{h_
   3^H}^\dagger\right)\Big]+4m_Z^4 {\rm Re}\left(h_1^H{h_
   3^H}^\dagger\right)-Q^4 {\rm Re}\left(h_2^H{h_
   3^H}^\dagger\right),
\end{align}
and  the $G\left(Q^2\right)$ function defined in Eq. \eqref{Gfunc}. From Eq. \eqref{Kfunction}, it is clear that the $\mathcal{A}_\phi$ asymmetry is very sensitive to the $CP$-violating form factor $h_3^H$.   We show in Fig. \ref{plotAsyAz} the $\mathcal{A}_\phi$ asymmetry as a function of the four-lepton invariant mass for the three new physics scenarios, whereas the SM case is not considered as it is zero. Once again, the largest values are reached for Scenario 1, which are order $10^{-4}$. For scenario 3, we obtain smaller results, since they are of order $10^{-6}$.  Contrary to the behavior observed for the $\mathcal{A}_{FB}$ asymmetry, we find that the $\mathcal{A}_\phi$ does not vanish at  $Q> 1000$ GeV. Thus, the azimuthal asymmetry may be measured at super high energies.

  The $\phi$ distributions are also good channels for observing the effects of anomalous couplings. Our SM results are in agreement with those reported in Ref. \cite{Godbole:2007cn}, nonetheless, the anomalous couplings contributions are very different, as the shift was not observed, and therefore the azimuthal asymmetry has not been studied in the form presented in this work.

In brief, the azimuthal distributions are sensitive to new physics, particularly to $CP$-violating effects. They can provide deeper insights into the potential impact on observables at high-energy colliders.

\section{Conclusions and outlook}
\label{conclusions}

We have presented a novel analysis of the $H^\ast\rightarrow ZZ \rightarrow \overline{\ell}_1\ell_1\overline{\ell}_2\ell_2$ decay width for both unpolarized and polarized $Z$ gauge bosons considering the most general $H^*ZZ$ vertex function, which is parametrized by three form factors $h^H_i$ ($i$=1, 2, 3), where the SM contributions up to the one-loop level and anomalous contributions arising from new-physics are included. We then consider the scenario where all the $h_i^H$  couplings are complex, which to our knowledge has never been studied in the literature, and obtain analytical results 
for both the unpolarized and polarized $H^\ast\rightarrow ZZ \rightarrow \overline{\ell}_1\ell_1\overline{\ell}_2\ell_2$  square amplitudes, out of which the decay widths,  angular distributions, and left-right (forward-backward) asymmetries can be straightforwardly obtained. Our results reproduce previous ones obtained in more restrictive scenarios.  
It is found that the polarized $H^\ast\rightarrow ZZ \rightarrow \overline{\ell}_1\ell_1\overline{\ell}_2\ell_2$ decay is mainly determined by the $H^\ast\rightarrow ZZ$ one for the four-lepton invariant mass distributions.

To cross-check the consistency of our numerical evaluation method we use \texttt{MadGraph5\_aMC@NLO}, with the corresponding Feynman rules for our model obtained via the FeynRules package. For the numerical analysis we consider some realistic scenarios for the complex $HZZ$ anomalous couplings, consistent with the current experimental and indirect constraints, and analyze the role of the transversal (left-handed and right-handed) polarizations of the $Z$ gauge bosons on the behavior of the $H^\ast\rightarrow ZZ \rightarrow \overline{\ell}_1\ell_1\overline{\ell}_2\ell_2$ decay width as a function of the four-lepton invariant mass $Q$, which turns out to be very distinctive for each type of polarization.  We found that for some specific values of the complex $HZZ$ couplings, the polarized $H^\ast\rightarrow ZZ \rightarrow \overline{\ell}_1\ell_1\overline{\ell}_2\ell_2$ decay width can deviate considerably from the SM contribution, but such a deviation is more pronounced at large $Q$. An opposite behavior between the left and right-handed polarizations is observed. This leads to a left-right asymmetry $\mathcal{A}_{LR}$, which vanishes in the SM as requires the presence of $CP$-violating anomalous couplings. 
The  angular distributions of the $H^\ast\rightarrow ZZ \rightarrow \overline{\ell}_1\ell_1\overline{\ell}_2\ell_2$ decay are also analyzed, which exhibits  similar behavior to that observed for the invariant mass distributions. Hence, we introduce an angular left-right  asymmetry $\mathcal{A}_{LR\theta}$ that has a non-vanishing SM one-loop level contribution but shows considerable deviations due to complex  anomalous $HZZ$ couplings. Finally, a polarized forward-backward asymmetry $\mathcal{A}^\lambda_{FB}$ is studied, which is found to be  constant but differs by a sign for left-handed and right-handed polarizations of the $Z$ gauge bosons. 
It is also worth noting that the $H^\ast\rightarrow ZZ \rightarrow \overline{\ell}_1\ell_1\overline{\ell}_2\ell_2$ decay is unsusceptible to new-physics effects in the case of longitudinally polarized $Z$ gauge bosons.
For completeness, we also study the scenario with unpolarized $Z$ gauge bosons, where the forward-backward  $\mathcal{A}^\lambda_{FB}$ and azimuthal $\mathcal{A}_\phi$ asymmetries as a function of the four-lepton invariant mass are examined, where a different behavior compared with the $\mathcal{A}_{LR}$ asymmetry is found.   

In summary, the study of the off-shell $H^\ast\rightarrow ZZ \rightarrow \overline{\ell}_1\ell_1\overline{\ell}_2\ell_2$ decay, via the transverse polarizations of the $Z$ gauge bosons, could be a useful tool to search for effects of new physics due to the real and absorptive parts of the $HZZ$ anomalous couplings, which could lead to significant deviations from the SM contributions in the decay widths as well as other observables such as angular distributions and left-right (forward-backward, azimuthal) asymmetries. In particular, we have put special emphasis on the study  of the effects of the absorptive parts of the $HZZ$ anomalous couplings as they have been largely overlooked in the past  but can serve as a probe of the SM at the LHC and future colliders. Such absorptive parts are an unique prediction of quantum field theory.

\begin{acknowledgments}
This work  was supported by UNAM Posdoctoral Program (POSDOC). We also acknowledge support from  Sistema Nacional de Investigadores (Mexico). 
\end{acknowledgments}

\appendix

\section{Kinematics and phase space for the $H\to ZZ\to \bar{\ell}_1\ell_1\bar{\ell}_2\ell_2$ decay}\label{kinematics}
The phase space of an $n$-body decay  was  studied  by Cabibbo and Maksymowicz \cite{Cabibbo:1965zzb}, Pais and Treiman \cite{Pais:1968zza}, and  Byckling and Kajanate \cite{Byckling:1969hni,Byckling:1969luw,Byckling:1969sx,Byckling:1969zz,Byckling:1971vca}. In particular, the $1 \rightarrow 4$ decay can be decomposed through a recursion relation into  3 partial $1\rightarrow 2$ decays  as follows \cite{Cabibbo:1965zzb,Byckling:1969sx,Cheng:1993ah}: $1\rightarrow 2 \longrightarrow  1\rightarrow 2 +1\rightarrow 2 $. 
Following this approach we obtain below the differential phase space for the $H\to ZZ\to \bar{\ell}_1\ell_1\bar{\ell}_2\ell_2$ decay.

For the four-momenta we use the nomenclature defined in Fig. \ref{KinFig}, namely,  $H(q)\to Z(p_1)+Z(p_2)$, followed by $Z(p_1)\to \ell_1(q_1)+\bar{\ell}_1(q_2)$ and $Z(p_2)\to\ell_2(q_3)+\bar{\ell}_2(q_4)$. For our calculation we consider the reference systems shown in Fig. \ref{plane}. The differential decay width is thus given in terms of the following three angles $\theta_1$, $\theta_2$ and $\phi$ angles:
\begin{itemize}
  \item $\theta_1$ is the angle between $\vec{p}_1$ and $\vec{q}_1$ in the rest frame of the $Z(p_1)$ gauge boson.
  \item $\theta_2$ is the angle between $\vec{p}_2$ and $\vec{q}_3$ in the rest frame of the $Z(p_2)$ gauge boson.
  \item $\phi$ is the relative angle between the $Z(p_1)\rightarrow 2\ell_1$ and $Z(p_2)\rightarrow 2\ell_2$ decay planes, taken as positive from the $Z(p_1)$ plane to the $Z(p_2)$ plane, with $\phi=0$ in the case in which both planes coincide  and both  $\vec{q_1}$ and $\vec{q_3}$ are in the same direction.
\end{itemize}

We now discuss the kinematics for the distinct reference frames introduced above.
\subsection{Off-shell Higgs boson rest frame}
  The kinematics of the  $H^\ast\rightarrow ZZ$ decay in the Higgs boson rest frame can be described by the following relations
\begin{align}
    & q^\mu=\big(Q,\ 0  \big),  \\
    &  p_{1,2}^\mu=\big(Q/2,\pm \vec{p}\big),
\end{align}
  where the magnitude of the three-momentum $\vec{p}$ is 
\begin{equation}
\|\vec{p}\|=\frac{\sqrt{Q^2-4 m_Z^2}}{2}.
\end{equation}
We consider that in this frame the $Z$ gauge bosons move along the $x$ axis, so their polarization vectors can be written as
\begin{align}
  & \epsilon^\mu_{1,2}(0)=\frac{1}{m_Z}\Big( \| \vec{p} \|, \pm \frac{\sqrt{Q^2}}{2} ,0  , 0  \Big), \\
    &  \epsilon^\mu_{1}(R/L)=\frac{1}{\sqrt{2}}\Big(0 , 0 , - i  , \pm1  \Big). \label{polRLvec}
\end{align}

\subsection{$Z(p_i)$ gauge boson rest frame}

In our calculation, we also use the rest frame of a $Z(p_i)$ gauge boson decaying into a lepton-antilepton pair, which obeys the following kinematics
\begin{align}
& p_i^\mu=\big(m_Z,\ 0  \big),  \\
    &  q_1^\mu=\big(m_Z/2,\ \vec{k}_1\big), \quad q_2^\mu=\big(m_Z/2,\ -\vec{k}_1\big),\\
    &  q_3^\mu=\big(m_Z/2,\ \vec{k}_2\big), \quad q_4^\mu=\big(m_Z/2,\ -\vec{k}_2\big),
\end{align}
where, according to Fig. \ref{plane},  the three-momenta $\vec{k}_i$ are given as
\begin{align}
    & \vec{k}_1=\frac{\sqrt{m_Z^2-4 m_{\ell_i}^2}}{2}\big(\cos{\theta_1} , \sin{\theta_1}, 0\big),  \\
    & \vec{k}_2=\frac{\sqrt{m_Z^2-4 m_{\ell_i}^2}}{2} \big(-\cos{\theta_2},\sin{\theta_2}\cos{\phi} , \sin{\theta_2}\sin{\phi}\big),
\end{align}
whereas the form of the transverse polarization vectors of Eq. \eqref{polRLvec} still holds, whereas for longitudinal polarizations we have
\begin{align}
   \epsilon^\mu_{1,2}(0)=\Big(0, \pm 1 ,0  , 0  \Big).
\end{align}

Since the interference term of the $H\to ZZ\to \bar{\ell}_1\ell_1\bar{\ell}_2\ell_2$ square amplitude has to be calculated in the Higgs boson frame, we need to boost the four-momenta and polarization vectors defined in the $Z(p_i)$ rest frames into the Higgs boson rest frame, which is achieved via the following Lorentz matrix 
\begin{equation}
\label{boost}
\Lambda^\mu_\nu=\left(\begin{array}{cccc}\gamma & \gamma v & 0 & 0 \\\gamma v & \gamma & 0 & 0 \\0 & 0 & 1 & 0 \\0 & 0 & 0 & 1\end{array}\right),
\end{equation}
with
\begin{equation}
v=\frac{\vec{p}_i}{E_i}=\pm \frac{\sqrt{Q^2-4m_Z^2}}{Q}.
\end{equation}

\subsection{Phase space}

The  $H\to ZZ\to \bar{\ell}_1\ell_1\bar{\ell}_2\ell_2$  phase space $d\left(R_4\right)$ can be written as 
\begin{align}
   d\left(R_4\right)= \frac{d^3q_1}{(2\pi)^3 2E_{q_1}}\frac{d^3q_2}{(2\pi)^3 2E_{q_2}}\frac{d^3q_3}{(2\pi)^3 2E_{q_3}}\frac{d^3q_4}{(2\pi)^3 2E_{q_4}}(2\pi)^4\delta^4\big(q-q_1-q_2-q_3-q_4\big),
\end{align}
where the lepton energies are $E_{q_i}=\sqrt{m_i^2+\!\vec{p}_i\!^2}$ ($i=1,2,3,4$). We will follow the  approach of Ref. \cite{Cheng:1993ah}, which is analogous to that used in \cite{Cabibbo:1965zzb}. We thus introduce the relations  
\begin{equation}
\label{fac1}
\int \frac{d^3 p_1}{2E_{p_1}} dS_1 \delta^4\big(p_1-q_1-q_2\big)=1,
\end{equation}    
and
\begin{equation}
\label{fac2}
\int \frac{d^3 p_2}{2E_{p_2}} dS_2 \delta^4\big(p_2-q_3-q_4\big)=1,
\end{equation}
where $E_{p_i}=\sqrt{{\vec{p}_i}^{\ 2}+S_i}$, with  $S_i=p_i^2$  ($i=1,2$).   

The $d\left(R_4\right)$ phase space can thus be written as
\begin{equation}
d\left(R_4\right)= \frac{dS_1 dS_2 }{(2\pi)^8}I_{p_1}I_{p_2}I_{q},
\end{equation}
where, in the rest frames of the $Z$ gauge bosons and the Higgs boson, the $I_{p_i}$  and $I_{q}$ integrals are given  for massless leptons as follows
\begin{align}
\label{int1}
     I_{p_1}&=\int \frac{d^3q_1}{2E_{q_1}} \frac{d^3q_2}{2E_{q_2}}\delta^4\big(p_1-q_1-q_2\big)\\
   &=\frac{\pi}{4}d\cos{\theta_1},
\end{align}
\begin{align}
\label{int2}
     I_{p_2}&=\int \frac{d^3q_3}{2E_{q_3}} \frac{d^3q_4}{2E_{q_4}}\delta^4\big(p_2-q_3-q_4\big)   \\
    &=\frac{1}{8}d\cos{\theta_2}d\phi,
\end{align}
and
\begin{align}
\label{int3}
     I_{q}&=\int \frac{d^3p_1}{2E_{p_1}} \frac{d^3p_2}{2E_{p_2}}\delta^4\big(q-p_1-p_2\big)   \\
    &=\frac{\pi}{2 Q}\sqrt{Q^2-4 m_Z^2}.    
\end{align}
Therefore the $d\left(R_4\right)$ phase space reads
\begin{align}
d\left(R_4\right)= \frac{\sqrt{Q^2-4 m_Z^2}}{256 Q (2\pi)^6} dS_1 dS_2 d\cos{\theta_1} d\cos{\theta_2}d\phi,
\end{align}

For massless leptons, the integration region reads
  \begin{align}
0< &\ S_2<\left(Q-\sqrt{S_1}\right)^2,\\
0< &\ S_1<Q^2,\\
0<&\ \theta_1,\ \theta_2 < \pi,\\
0<&\ \phi<2\pi,
\end{align} 
where $Q^2=q^2$ is the invariant mass of the four final leptons,which is the same in all frames of reference and is usually used in LHC analyses. For the purpose of our calculation we denote $Q=\|q\|$.

\bibliography{BiblioH}

\end{document}